\documentclass[aps,pra,twocolumn,superscriptaddress]{revtex4-2}

\usepackage[utf8]{inputenc}
\usepackage{amsmath,amssymb,amsfonts}
\usepackage{graphicx}
\usepackage{siunitx}
\usepackage{multirow}
\usepackage{booktabs}

\usepackage{subfigure}
\usepackage{bm} 
\usepackage{xcolor}

\newcommand{\eref}[1]{Eq.\,(\ref{#1})}
\newcommand{\fref}[1]{Fig.\,\ref{#1}}
\newcommand{\tref}[1]{Table\,\ref{#1}}
\newcommand{\sref}[1]{Section\,\ref{#1}}

\usepackage{braket}	

\newcommand{\ThreeJ}[6]{\left( \begin{matrix}  #1 & #2 & #3 \\ #4 & #5 & #6 \end{matrix} \right)}
\newcommand{\SixJ}[6]{\left \{ \begin{matrix}  #1 & #2 & #3 \\ #4 & #5 & #6 \end{matrix} \right \} }

\newcommand*{\wn}{cm$^{-1}$}

\usepackage[normalem]{ulem}

\begin{document}


\title{Precision millimetre-wave spectroscopy and calculation of the Stark manifolds in high Rydberg states of para-H$_2$}

\author{N. H\"{o}lsch}
\email{nicolas.hoelsch@phys.chem.ethz.ch}
\affiliation{Laboratorium f\"{u}r Physikalische Chemie, ETH Z\"{u}rich, 8093 Z\"{u}rich, Switzerland}

\author{I. Doran}
\affiliation{Laboratorium f\"{u}r Physikalische Chemie, ETH Z\"{u}rich, 8093 Z\"{u}rich, Switzerland}

\author{M. Beyer}
\affiliation{Department of Physics and Astronomy, LaserLaB, Vrije Universiteit Amsterdam, de Boelelaan 1081, 1081 HV Amsterdam, The Netherlands}

\author{F. Merkt}
\email{frederic.merkt@phys.chem.ethz.ch}
\affiliation{Laboratorium f\"{u}r Physikalische Chemie, ETH Z\"{u}rich, 8093 Z\"{u}rich, Switzerland}

\begin{abstract}
Precision measurements of transitions between singlet ($S=0$) Rydberg states of H$_2$ belonging to series converging on the $\mathrm{X}^+\,^2\Sigma_g^+(v^+=0,N^+=0)$ state of H$_2^+$ have been carried out by millimetre-wave spectroscopy under field-free conditions and in the presence of weak static electric fields. The Stark effect mixes states with different values of the orbital-angular-momentum quantum number $\ell$ and leads to quadratic Stark shifts of low-$\ell$ states and to linear Stark shifts of the nearly degenerate manifold of high-$\ell$ states. Transitions to the Stark manifold were observed for the principal numbers 50 and 70, at fields below 50 mV/cm, with linewidths below 500~kHz.
The energy-level structure was calculated using a matrix-diagonalisation approach, in which the zero-field positions of the $\ell\leq 3$ Rydberg states were obtained either from multichannel-quantum-defect-theory calculations or experiment, and those of the $\ell\geq 4$ Rydberg states from a long-range core-polarisation model. This approach offers the advantage of including rovibronic channel interactions through the MQDT treatment while retaining the advantages of a spherical basis for the determination of the off-diagonal elements of the Stark operator. 
Comparison of experimental and calculated transition frequencies enabled the quantitative description of the Stark manifolds, with residuals typically below 50 kHz. We demonstrate how the procedure leads to quantum defects and binding energies of high Rydberg states with unprecedented accuracy, opening up new prospects for the determination of ionisation energies in molecules.
\end{abstract}

\maketitle



%
%
%



\section{Introduction}
\label{sec:intro}

Rydberg states of atoms and molecules are very sensitive to electric
fields \cite{stebbings83a, gallagher94a}. At small fields, nondegenerate states of low
orbital-angular-momentum quantum number $\ell$ undergo quadratic shifts and
their polarisability scales to a first approximation with the seventh power of the principal
quantum number $n$. Core-nonpenetrating high-$\ell$ states exhibit
linear Stark shifts already at low electric fields and the shifts are
very large, scaling with $n^2$ for the most blue- and red-shifted states
of the Stark manifolds. These properties make Rydberg Stark states ideal
electric-field sensors \cite{osterwalder99a,facon16a,koepsell17a} and
form the basis for efficient methods to control the translational motion
of Rydberg atoms and molecules with inhomogeneous electric fields \cite{vliegen04a, seiler11a, hogan16a}. 

At the same time, the large Stark  shifts of high Rydberg states are
detrimental in precision measurements of the spectral positions of
Rydberg states, as used, for instance in the determination of ionisation
energies and quantum defects in atoms
\cite{mack11a, sassmannshausen13a, clausen21a} and molecules
\cite{osterwalder04a, beyer18a}. For these reasons, accurate
measurements and calculations of the Stark effect in high atomic and
molecular Rydberg states are required. Two strategies can be followed in
precision measurements of high Rydberg states: The first one consists of
exploiting electric-field magic transitions \cite{peper19b}, i.e.,
transitions that are insensitive to electric fields, because the Stark
shifts of the initial and final states are identical. The second is to
quantitatively account for the Stark shifts in well-defined electric
fields. Both methods require the ability to accurately compute the
spectra of Rydberg atoms and molecules in the presence of electric fields.

The advent of narrow-band tunable lasers in the 1970s enabled the first high-resolution spectroscopic studies of the Stark effect in Rydberg states of the alkali-metal atoms \cite{littman76a}. The Stark spectra revealed complex spectral structures arising from the coupling, through the electric field, of series of different $\ell$ values. An efficient procedure to calculate the Stark effect was introduced by Zimmerman et al. \cite{zimmerman79a}, relying on a matrix representation of the Hamiltonian. Today, programs are available to precisely compute Stark spectra of alkali-metal atoms \cite{vsibalic17a,weber17a}. In $^{87}$Rb, this treatment has been verified experimentally to an accuracy of up to 2~MHz for $n=70$ and fields around 6~V/cm \cite{grimmel15a}.

The treatment of the Stark effect in Rydberg atoms and molecules with
open-shell ion cores, such as the rare-gas atoms and
molecular hydrogen, is more challenging, because the channel
interactions between Rydberg series converging on different quantum
states of the ion lead to perturbations that are not described by
Rydberg's formula. Matrix-diagonalisation methods have been developed
to account for these channel interactions and also describe
autoionisation resonances \cite{ernst88a,fielding91b}. Approaches based
on multichannel quantum defect theory (MQDT), however, more naturally and
elegantly include the channel interactions
\cite{harmin81a,sakimoto86a,sakimoto89a}. Such approaches divide the
configuration space in three regions, an inner, close-coupling region
where the channel interactions originate, an intermediate region, where
the effects of the Coulomb potential are dominant, and a long-range
region, where the effects of both the Coulomb interaction and the
electric field need to be considered. Calculations involve two frame
transformations. The first one, between the first and second regions,
corresponds to the usual frame transformation of MQDT
\cite{seaton83a,jungen11a}. The second one connects the generalised
Coulomb functions describing the intermediate region to the parabolic
functions with which Rydberg-Stark states are best described in the
outer region. Such approaches have been used by Softley and coworkers to
analyse Stark spectra of Ar \cite{fielding92a}, H$_2$ \cite{fielding94a}
and Ne \cite{gruetter08a} Rydberg states. Although overall very good
agreement between experimental and calculated spectra could be reached,
line positions and intensities did not always perfectly match,
indicating either possible deficiencies in the quantum defects used or a
nonnegligible role of the electric field in the second region.

In this article, we present an experimental and computational study of
the Stark effect in high Rydberg states of H$_2$. The emphasis is placed
on the accurate treatment of the Stark effect in Rydberg
states of principal quantum number in the range between 40 and 80, a region that was used in our previous determination of the ionisation energy of para-H$_2$ using Rydberg-series extrapolation \cite{beyer19a}. We
aim at reaching a level of precision better than 50 kHz required by the
next generation of ionisation-energy determination in H$_2$ \cite{hoelsch19a}. In order to reach this level of accuracy, we
record high-resolution millimetre-wave spectra of Rydberg-Rydberg
transitions in weak, well-controlled electric fields and compare the
experimental spectra with spectra calculated numerically in an approach
combining the diagonalisation of the Stark Hamiltonian matrix with MQDT for the description of penetrating low-$\ell$ Rydberg states at zero-field.

Molecular hydrogen is an ideal molecular system to test this approach.
Firstly, hydrogen is the simplest two-electron molecule and
precision determination of its dissociation and ionisation energies play
an important role in testing fundamental theories of molecular structure
that include nonadiabatic, relativistic and QED corrections
\cite{wolniewicz95c,piszczatowski09a,puchalski19a}.
Secondly, the treatment of both penetrating and nonpenetrating Rydberg
states of H$_2$, HD and D$_2$ by MQDT is extremely advanced thanks to
systematic progress over several decades by Jungen and his coworkers
\cite{herzberg72a,jungen77a,jungen84a,jungen97a,jungen98a}.
Thirdly, previous high-resolution spectroscopic studies of penetrating
\cite{osterwalder00a,osterwalder04a} and
nonpenetrating \cite{eyler83a,jungen89a,jungen90a,uy00a,lundeen05a}
Rydberg states of H$_2$ have been reported, which form an excellent
starting point for further precision studies. Finally, previous work on
high-$n$ states of H$_2$ has demonstrated that nonpenetrating states
offer significant advantages for the determination of ionisation
energies by Rydberg-series extrapolation: they have small quantum
defects and are much less perturbed by channel
interactions than low-$\ell$ states \cite{beyer19a}. 

The large sensitivities to electric fields of nonpenetrating high-$\ell$ states are a disadvantage in zero-field measurements because of uncontrollable shifts caused by stray electric fields, but can be exploited to our advantage in Stark measurements. We show in this article, that the combination of a 3-photon excitation scheme with application of weak electric fields provides easy optical access to the full range of Rydberg Stark states. For high Rydberg states, even weak electric fields of the order of a few 10~mV/cm lead to all $\ell>3$ states being merged into a manifold with almost linear shifts. The states of this Stark manifold are desirable as spectroscopic targets because their positions are less sensitive to uncertainties in the quantum defects, a limiting factor in the determination of ionisation energies by Rydberg series extrapolation. In turn, the high-$\ell$ Stark manifold constitutes a reference for the positions of the respective $\ell=3$ states and therefore provides a way to assess the MQDT calculations at the highest precision.

\section{Experimental Methods}
\label{sec:methods}

Millimetre-wave (mmW) spectra of transitions between high-$n$ Rydberg states of para-H$_2$ were recorded at high spectral resolution in the presence of electric fields using a resonant three-photon excitation scheme as described in previous studies of high-$n$ p, d and f Rydberg states of ortho-H$_2$ \cite{osterwalder00a, osterwalder04a}. Whereas the earlier work was devoted to the characterisation of the hyperfine structure of high-$n$ p, d, and f Rydberg states belonging to series converging on the $\mathrm{X}^+\,^2\Sigma_g^+(v^+=0,N^+=1)$ ground state of ortho-H$_2^+$, the present study focusses on the Stark effect in Rydberg series converging on the $\mathrm{X}^+(v^+=0,N^+=0)$ ground state of para-H$_2^+$. We describe here in detail the aspects of the measurements related to the study of the Stark effect and only briefly recapitulate the main aspects of the experimental setup and methods, which are presented in detail in Ref.~\cite{osterwalder04a}. In order to label Rydberg states, we use the notation $n\ell N^+_N$, where $N$ is the quantum number of the total angular momentum without spin, $\vec{N}=\vec{N}^++\vec{\ell}$, and therefore $\vec{N}=\vec{\ell}$ in the case of $\vec{N}^+=0$.

\begin{figure}[ht]
	\centering
	\includegraphics[trim=0cm 0cm 0cm 0cm, clip=true, width=0.95\linewidth]{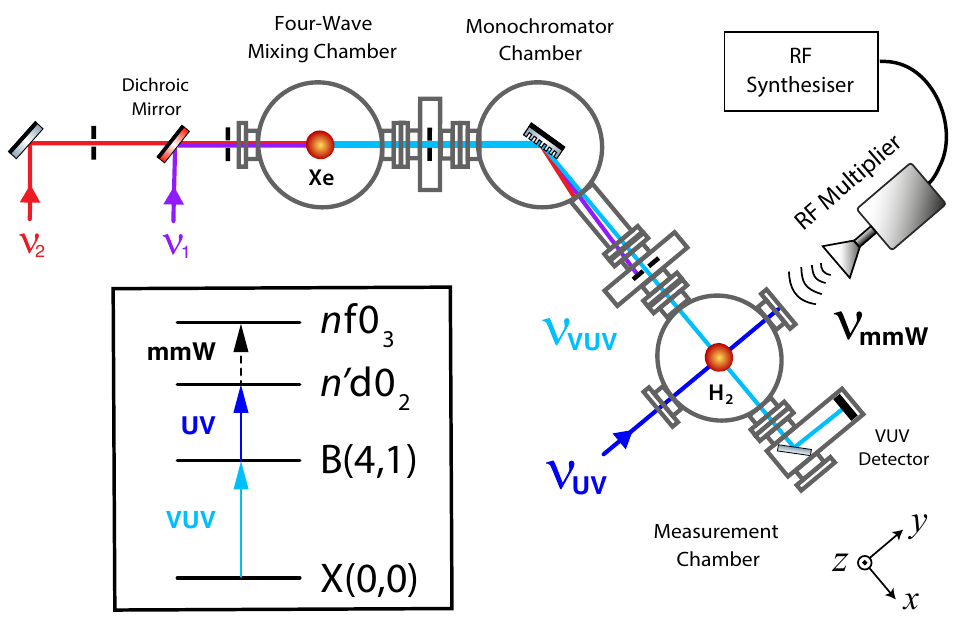}
	\caption{Schematic representation of the experimental setup. The molecular beams travel perpendicular to the plane of the lasers and are indicated by red dots. The three-photon excitation sequence used to reach the high Rydberg states of H$_2$ is presented in the inset. See text for details.}
	\label{fig:setup}
\end{figure}

\begin{figure}[ht]
	\centering
	\includegraphics[width=\linewidth]{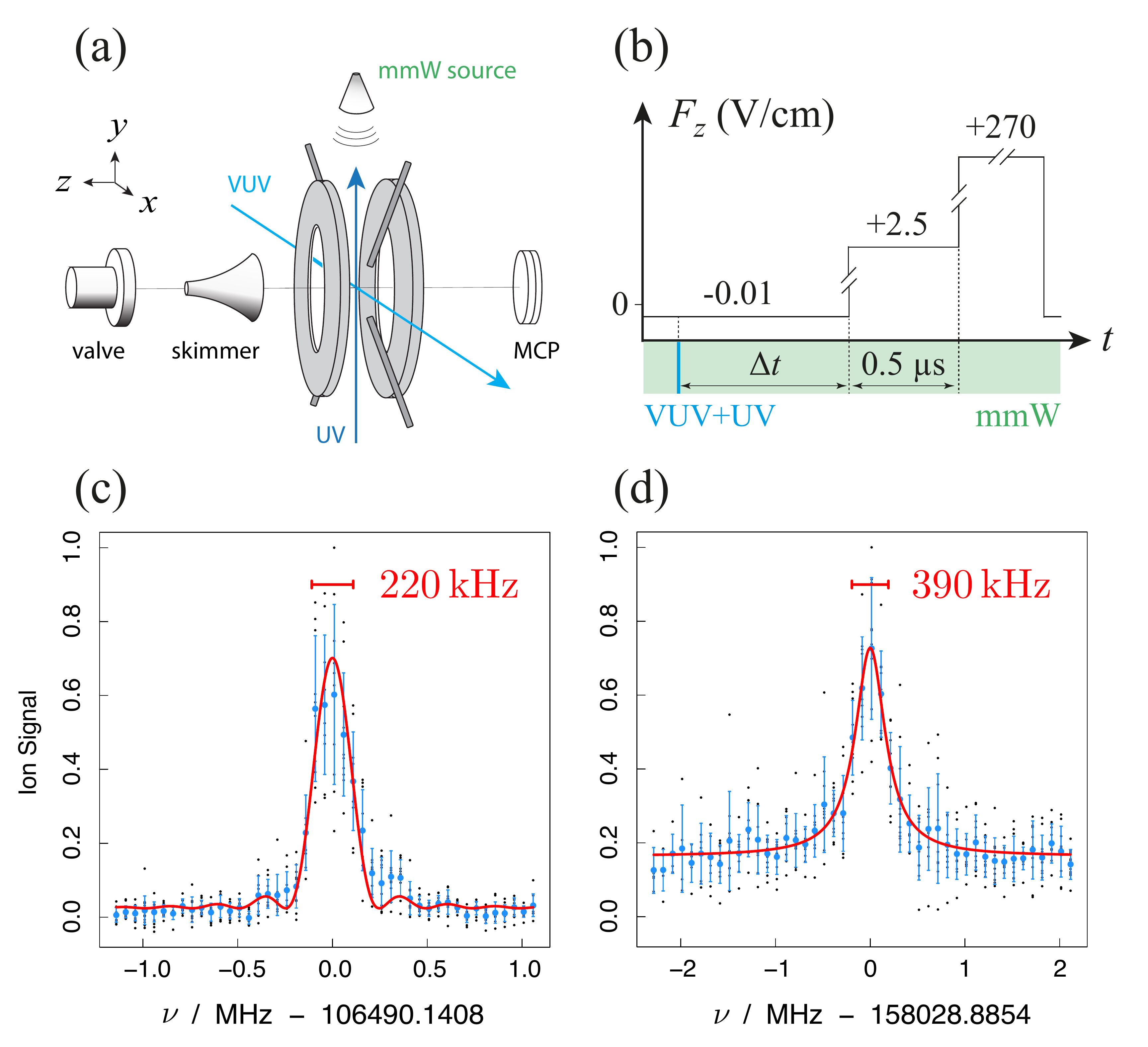}
	\caption{
	(a) Schematic layout of the electrodes around the photoexcitation region. The molecular beam is indicated as a horizontal line and the intersecting laser beams as coloured arrows. Only two of the six ring electrodes of the electrode stack are shown. Four electrode rods penetrate the stack in the plane of the laser beams. (b) Example of a temporal sequence of electric fields used to detect mmW Rydberg-Rydberg transitions in H$_2$. The presence of laser and mmW radiation is indicated in colour. (c) and (d) Spectra of the 51f$0_3\leftarrow 49$d$0_2$ and 70f$0_3\leftarrow 63$d$0_2$ transitions of H$_2$, respectively, recorded for a measurement time $\Delta t$ of $4\,\mu\mathrm{s}$. The black dots represent the H$_2^+$ ion signal observed in individual measurements. Each blue dot with vertical error bar correspond to the average and standard deviation of six measurements, and the red lines are fits of the line profiles using a sinc$^2$ function and a Lorentzian, respectively (see text for details).
	\label{fig:measurement}
	}
\end{figure}

The experimental setup is depicted schematically in Fig.~\ref{fig:setup}. A pulsed dye-laser system was used to excite selected members of the $n$d$0_2$ series converging on the $\mathrm{X}^+(0,0)$ ground state of para-H$_2^+$ from the $\mathrm{X}\,^1\Sigma_g^+(v=0,N=0)$ ground state of H$_2$ using a resonant two-photon excitation process via the $\mathrm{B}\,^1\Sigma_u^+(v=4,N=1)$ intermediate state (see inset of Fig.~\ref{fig:setup}). The pulsed VUV radiation ($\tilde{\nu}_\mathrm{VUV} = 2\tilde{\nu}_1+\tilde{\nu}_2$) used for the B$\leftarrow$X transition was produced by resonance-enhanced four-wave mixing in xenon using two pulsed lasers of wavenumbers $\tilde{\nu}_1$ and $\tilde{\nu}_2$, exploiting the $(5{\rm p})^5(6p)[1/2]_0\leftarrow (5{\rm p})^6\ ^1{\rm S}_0$ two-photon resonance of Xe at $2\tilde{\nu}_1=80118.97$~cm$^{-1}$. $n$d Rydberg states were then excited using the frequency-doubled output of a third pulsed dye-laser, $\tilde{\nu}_\mathrm{UV} \approx 29100\,$\wn. This laser pulse was temporally and spatially overlapped with the VUV laser pulse in the excitation chamber, crossing a skimmed supersonic beam of H$_2$ at right angles, see Figs.\,\ref{fig:setup} and \ref{fig:measurement}a. The supersonic beam was generated from a pure natural sample of H$_2$ using a cryogenic pulsed valve held at a temperature $T_{\rm valve}= 80$ K, which enabled the generation of a dense beam with central velocity of about 1400~m/s.

Transitions from the selected $n$d$0_2$ Rydberg states to higher-lying Rydberg states were recorded using narrow-band continuous-wave mmW radiation. The general procedure is described in Ref.~\cite{merkt98b} and the mmW source, tunable in the range 110-170~GHz, was presented in Ref.~\cite{peper19a}. The output of a radio-frequency (rf) synthesiser (\textsc{Agilent E8257D}) was used after harmonic generation using an active 12-fold multiplier (110-170~GHz, \textsc{Virginia Diodes WR-6.5}). The mmW radiation passed a calibrated, adjustable attenuator before being coupled to free-space by a horn and sent into the vacuum chamber through an optical viewport. Sheets of paper were added in front of the chamber when additional attenuation was required. The rf synthesiser was referenced to a GPS-disciplined rubidium atomic clock (\textsc{Stanford Research Systems FS725} with a \textsc{Spectrum Instruments TM-4} GPS receiver).

The mmW spectra were recorded by monitoring the H$_2^+$ ions generated by delayed pulsed field ionisation as a function of the mmW frequency using the electric-field pulse sequence depicted in Fig.~\ref{fig:measurement}b. In this sequence, a two-step pulse is applied across a cylindrical electrode stack after an adjustable delay to selectively field ionise the upper Rydberg level of the transitions and extract the H$_2^+$ ions toward a microchannel-plate detector along a flight tube with axis oriented perpendicular to the plane of Fig.~\ref{fig:setup}. The measurement time for the mmW transitions corresponds to the interval $\Delta t$ between the laser pulse and the onset of the first electric-field step. This prepulse produces a field of several V/cm that is used to separate prompt ions produced by photoionisation from the neutral H$_2$ Rydberg molecules. This field is not strong enough to field ionise the Rydberg states involved in the transitions but large enough to shift the target states out of resonance. Consequently, Rydberg-Rydberg transitions are entirely suppressed at the end of the interval $\Delta t$. The second, stronger field step is then used to field ionise the upper Rydberg state of the transitions. Its strength is carefully adjusted (e.g., to $\approx 23.3$~V/cm at $n=70$) to avoid the field ionisation of the lower-lying, initial $n$d$0_2$ Rydberg states.

In addition to the pulsed potential applied across the electrode stack, static potentials are also used to compensate stray electric fields in the photoexcitation volume. The compensation is carried out by measuring, and then minimising, the quadratic Stark shifts of selected transitions (e.g., the $n$f$0_3\leftarrow n^\prime$d$0_2$ transitions), following the procedure described in Refs.~\cite{osterwalder99a,beyer18a}. The stray field components along, and perpendicular to, the direction of the flight-tube axis are adjusted by applying potentials to the electrodes of the extraction stack and to four lateral metallic rods, respectively. When measuring Stark spectra, well-defined potential differences are applied across the electrode stack while making sure that the potentials applied to the metallic rods compensate the transverse components of the stray electric field.

The static electric fields applied across the electrode stack define the quantisation axis, i.e., the $z$-axis in Fig.~\ref{fig:setup}. In all experiments, the polarisation of the VUV laser was held parallel to the quantisation axis, so that only the $M_N=0$ component of the B(4,1) intermediate state was accessed from the ground state. The polarisation of the UV laser light could be rotated to excite either the $|M_N|=0$ or the $|M_N|=1$ component of the initial $n$d$0_2$ Rydberg states. Stark spectra were recorded for both parallel and perpendicular UV-laser polarisation. The polarisation of the millimetre-wave radiation was predominantly parallel to the quantisation axis.

Doppler broadening of the mmW transitions is negligible under our experimental conditions. The line\-widths and line\-shapes thus result from the lifetimes or measurement times, whichever lead to the larger broadening. Examples are presented in Fig.~\ref{fig:measurement}c and Fig.~\ref{fig:measurement}d. In the case of the 51f$0_3\leftarrow 49$d$0_2$ transition (Fig.~\ref{fig:measurement}c), the lineshape corresponds to a sinc$^2$ function with full-width at half maximum of 220~kHz, as expected for a measurement time $\Delta t\approx4\mu$s. In the case of the 70f$0_3\leftarrow 63$d$0_2$ (Fig.~\ref{fig:measurement}d), a Lorentzian lineshape with full width at half maximum of 390 kHz is observed. Using a pump-probe measurement scheme as described in Ref.~\cite{hoelsch18a}, the lifetimes of the $49$d$0_2$ and $63$d$0_2$ were determined to be 4.1(8)~$\mu$s and 410(30)~ns, respectively. The resulting lifetime broadenings are 40~kHz and 380~kHz, respectively, which is in agreement with the observed lineshapes. In the Stark spectra, we did not observe any narrowing of the lines for interaction times longer than $\Delta t = 4\,\mu\mathrm{s}$, even in cases where the lifetimes of the initial states would have been long enough. We conclude that, beyond $4\,\mu\mathrm{s}$, the molecules reach regions where the stray electric field is no longer perfectly compensated, which leads to an inhomogeneous broadening through the Stark effect.

Typical full widths at half maximum of the transitions to the Rydberg-Stark states presented in Section~\ref{sec:results} are in the range between 300 and 600 kHz, with no significant asymmetry and no dependence on the Stark shift, which indicates that field inhomogeneities did not affect the line profiles. The electric fields in the photoexcitation region are proportional to the electric-potential difference applied across the electrode stack. To determine the proportionality constant, we first estimated it from numerical simulations of the field distributions using a finite-element programme, and then refined it in a global comparison of transitions frequencies measured for different electric potentials and at different $n$ values with frequencies calculated with the method presented in the next section (see Section~\ref{sec:fit} for the details of the comparison). The refined proportionality constant differed from the initially estimated one by less than 1\,\%. In the comparison, we also considered the possibility of a residual transverse stray field after compensation and found it to be less than 200~$\mu$V/cm for all measurements presented in Section~\ref{sec:results}.

\section{Calculations of Stark Maps }
\label{sec:calc}

The Hamiltonian of a Rydberg atom or molecule in an electric field can be written in the form
\begin{equation}
\hat{H} = \hat{H}_0 + \hat{H}_\mathrm{Stark} = \hat{H}_0 + eF\hat{z}\,.
\label{eq:hamiltonian}
\end{equation}
Its eigenstates can be obtained in a matrix-diagonalisation approach using a suitable basis for the matrix representation of the effective Hamiltonian \cite{zimmerman79a}. This method was already employed in previous studies of the Stark effect in H$_2$ \cite{fielding91b, yamakita04a, seiler11b}. As in these previous works, we chose a Hund's case (d) basis $\ket{n \ell N^+ N M_N}$ but our method differs from earlier work in the computation of the zero-field energies.

\subsection{Field-free Hamiltonian using MQDT}
\label{sec:Hzero}
\vspace*{5pt}

The accurate calculation of Stark-shifted energy levels at low to moderate field strengths depends critically on the quality of the zero-field energies. In any description of molecular Rydberg states, the effects of rovibrational series interactions have to be taken into account. This situation is more complicated than in alkali-metal atoms, for which the zero-field positions are accurately described by the extended Rydberg-Ritz formula \cite{ritz08a}.

The most precise description of the Rydberg states of H$_2$ is achieved by using multichannel quantum defect theory (MQDT) \cite{jungen11a}. In a first step, energy- and $R$-dependent quantum defects are obtained in Hund's case (b) in a clamped-nuclei picture from \emph{ab initio} potential curves or a core-polarisation model \cite{jungen89a, jungen90a}. This relatively small set of parameters can then be used to derive the effective quantum defects for any Rydberg state belonging to series converging to a state ($v^+,N^+$) of the ion via a rovibrational frame-transformation method \cite{jungen77a}.

Descriptions of the Stark effect in NO \cite{vrakking96a} and H$_2$ \cite{seiler11b} employed a simplified model, using only ten effective Hund's-case-(b) quantum defects $\mu_{\ell\Lambda}$ and a rotational frame-transformation to obtain the zero-field Hamiltonian matrix in a Hund's case (d) basis, in which the Stark Hamiltonian was computed. In the present work, we use a full MQDT calculation for Rydberg states with $\ell=(0,1,2,3)$ to obtain level energies for the Rydberg series in para-H$_2$ converging on the first ionisation threshold. We include all relevant rotational channels, and vibrational channels up to $v^+=8$. For $N^+=0$, no hyperfine structure is present and there is no singlet-triplet mixing. In our experiment, only singlet states can be accessed and consequently only singlet configurations were considered in our calculations. For the $n$p and $n$f states, we use the quantum defects reported by Sprecher et al. \cite{sprecher14b} and Osterwalder et al. \cite{osterwalder04a} who estimated the accuracy of the resulting level positions to be better than 1~MHz. The quantum defects for the $n$s and $n$d states were previously derived \cite{sprecherDiss} following the procedure introduced in Ref.~\cite{ross94a}. Owing to strong nonadiabatic interactions with doubly excited states and the relatively strong s-d interaction, the MQDT treatment of these \emph{gerade} states is more challenging than for the \emph{ungerade} states at small internuclear distances. Even though these quantum defects have been shown to reproduce measured binding energies of high-$n$ ($n=50-60$) Rydberg states to within about 1~MHz, this agreement was only confirmed for the $n$s$1_1$ and $n$d$1_1$ series in ortho-H$_2$ so far \cite{sprecherDiss}. 

\begin{figure}[ht]
	\centering
	\includegraphics[trim=0cm 0cm 0.5cm 0cm, clip=true, width=0.9\linewidth]{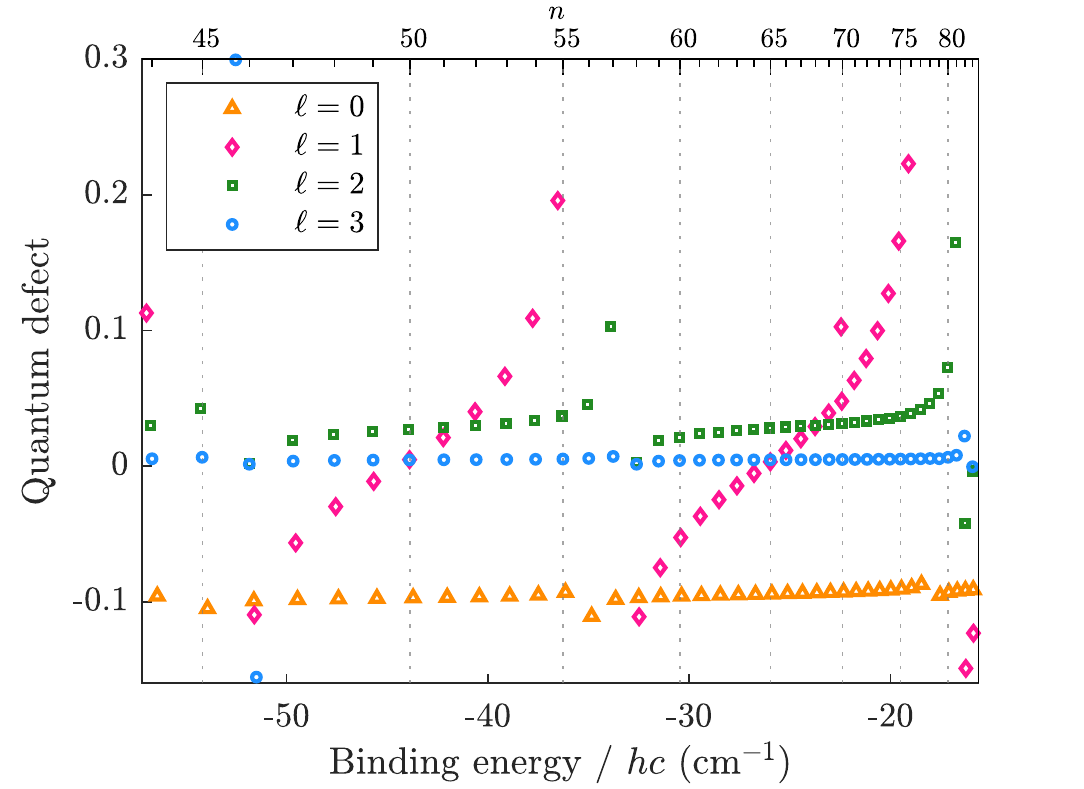}
	\caption{Effective quantum defects obtained from MQDT calculations (see text for details) for the members of the $n\mathrm{s}0_0$, $n\mathrm{p}0_1$, $n\mathrm{d}0_2$ and $n\mathrm{f}0_3$ series in the range $n=44-84$.}
	\label{fig:lufano}
\end{figure}

The effective quantum defects resulting from the calculated level positions in the region of interest are displayed in \fref{fig:lufano}. They follow a typical Lu-Fano-type behaviour characteristic of perturbed Rydberg series \cite{lu70a}. The perturbations around $n=45,57,80$ originate from rotational perturbers from series converging on the $\mathrm{X}^+(v^+=0,N^+=2)$ level of the ion. The strength of this rotational channel interaction decreases with increasing value of $\ell$ because the core penetration decreases. The $n$s$0_0$ series is the only series with $\ell=0,N=0$ and thus unperturbed to first approximation. The quadrupole moment of the core leads to s-d interaction with $n$d$2_0$ states. At $n=70$, an additional state with $\ell=1$ is predicted, which is assigned to an $n=4$ state belonging to a series converging on the $\mathrm{X}^+(v^+=4,N^+=0)$ state of H$^+_2$. However, this vibrational perturber appears to have only a small effect on the series converging on the ionic ground state, because the Lu-Fano behaviour of the quantum defects is not significantly perturbed.

Owing to the channel interactions, the Rydberg states have mixed character in the Hund's case (d) basis. In a full treatment, this mixed character needs to be taken into account in the evaluation of the matrix elements of the Stark Hamiltonian. The Rydberg states for which we measured mmW spectra were, however, not significantly perturbed by channel interactions. Consequently, we could calculate the matrix elements in the approximation that the eigenstates determined in the MQDT calculations are pure Hund's case (d) states. Because the effect of channel interactions on the position of the Rydberg states is taken into account in the zero-field energies, we can restrict our basis for the calculations to states with $N^+=0$. At the field strengths of less than 1~V/cm relevant for our study, the Stark effect does not affect the rovibrational state of the ion core and $N^+$ remains a good quantum number when an electric field is included.

For states with $\ell>3$, we calculate the quantum defects using the core-polarisation model derived by Eyler and Pipkin \cite{eyler83a}. In order to assess the quality of the obtained quantum defects, we compare them in the case of $\ell=4,5$ to quantum defects obtained from a more sophisticated core-polarisation model in combination with an MQDT treatment \cite{jungen89a, jungen90a, jungenPrivate}. We find the discrepancy between the two sets of quantum defects to be well below $\pm 10\%$ in regions where the rotational channel interactions with $N^+=2$ series are weak ($n=47-55$ and $n=60-75$). This discrepancy does not lead to a significant modification in the calculation of the observed high-$\ell$ states (see Section~\ref{sec:sensitivities}).

\subsection{Stark Hamiltonian Matrix}
\label{sec:Hstark}
\vspace*{5pt}

In the presence of an external field $\vec{F}=(0,0,F)$ chosen along the $z$-axis, the term $\hat{H}_\mathrm{Stark} = eF\hat{z}$ has to be added to the field-free Hamiltonian, see \eref{eq:hamiltonian}. The off-diagonal matrix elements of $\hat{H}_\mathrm{Stark}$ couple different field-free states and the eigenenergies in the field are obtained by diagonalisation. Using the Wigner-Eckart theorem (with $\hat{z}=r\cos\theta=T^{(1)}_0$) and standard angular-momentum algebra (Chapters 4.1 and 5.2 of \cite{zare88a}), the Stark matrix elements in our Hund's case (d) basis are evaluated using \cite{vrakking96a}
\begin{align}
\begin{split}
\braket{\psi | \hat{H}_\mathrm{Stark} | \psi^\prime} = &\braket{n \ell N^+ N M_N | eF\hat{z} | n' \ell' N^{+\prime} N' M'_N} \\ 
= & \delta _{N^+N^{+\prime}} \,eF\, (-1)^{N-M_N+\ell+N^+ +N'+1}\\
&\times \sqrt{(2N+1)(2N'+1)}\\
&\times \ThreeJ{N}{1}{N'}{-M_N}{0}{M'_N} \SixJ{\ell}{N}{N^+}{N'}{\ell'}{1} \\
&\times \braket{n\ell || r || n'\ell'}\,,
\end{split}
\label{eq:fullsh2}
\end{align}
where $\braket{n\ell || r || n'\ell'}$ is a reduced matrix element. This reduced matrix element is related to a radial integral via 
\begin{align}
\begin{split}
\braket{n\ell || r || n'\ell'} =& (-1)^{\ell}\sqrt{(2\ell+1)(2\ell'+1)} \\
&\times \ThreeJ{\ell}{1}{\ell'}{0}{0}{0}\braket{n\ell | r | n'\ell'}\,,
\label{eq:redmat}
\end{split}
\end{align}
using the Wigner-Eckart theorem and the Clebsch-Gordan series \cite{knight85b}. The properties of the Wigner 3-j and Wigner 6-j symbols in Eqs.~(\ref{eq:fullsh2}) and (\ref{eq:redmat}) result in the selection rules $\Delta M_N = 0$, $\Delta N^+ = 0$, $\Delta\ell = \pm 1$ and $\Delta N = 0, \pm 1 \,(0\nleftrightarrow0)$  for the Stark interaction. \\

\begin{figure*}[ht]
	\centering
	\subfigure{\includegraphics[trim=0cm 0cm 0cm 0cm, clip=true, width=0.3\linewidth]{
		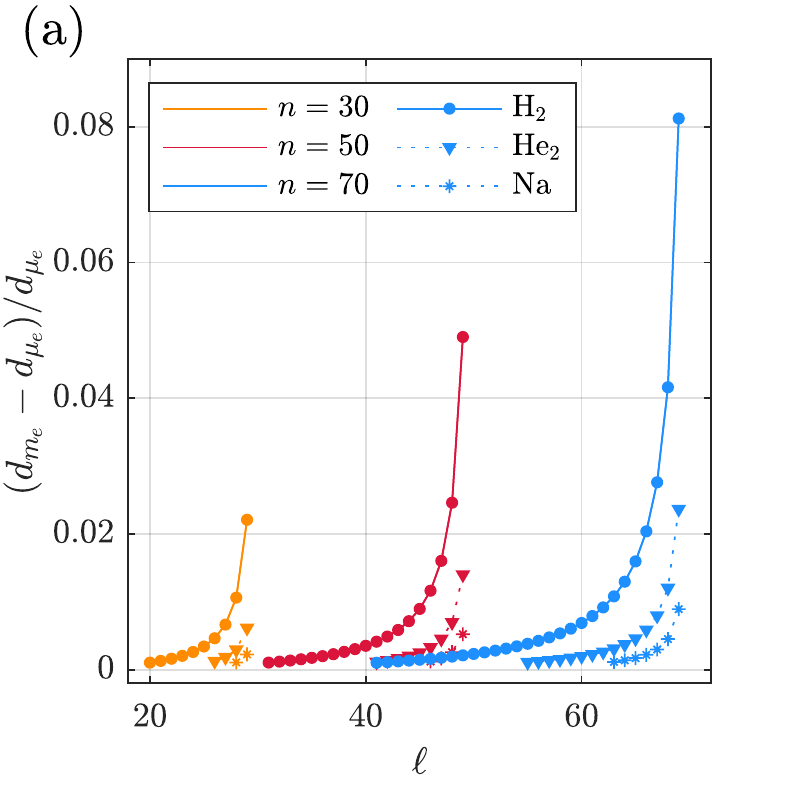}}
	\subfigure{\includegraphics[trim=0cm 0.3cm 0cm 0cm, clip=true, width=0.33\linewidth]{
		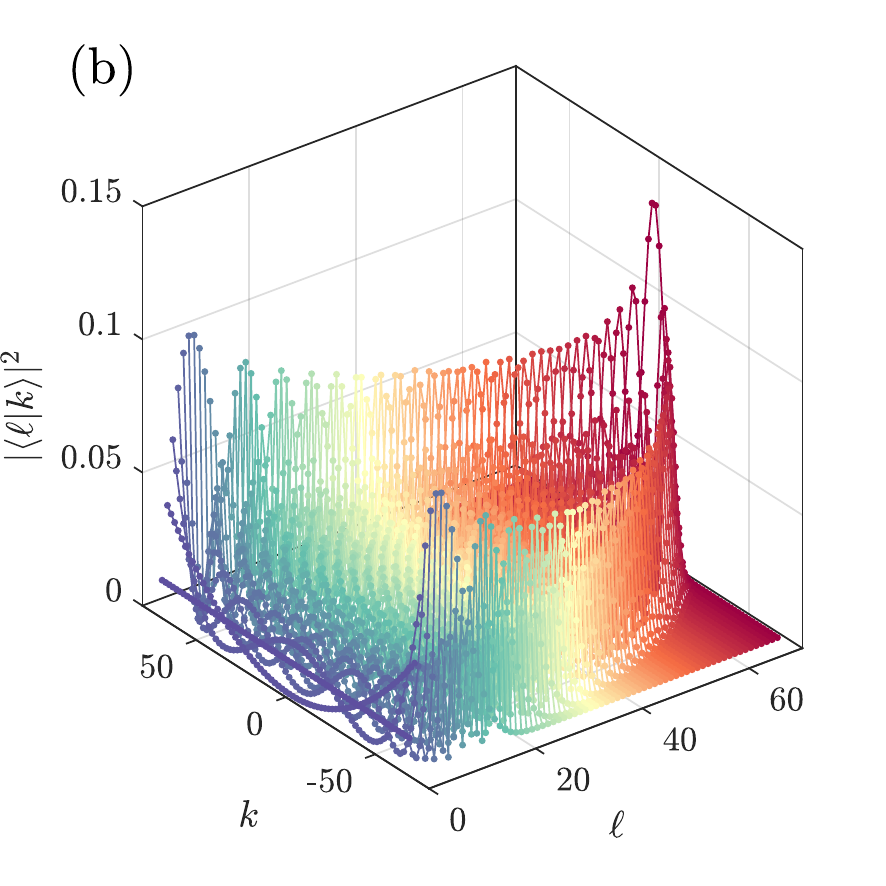}}
	\subfigure{\includegraphics[trim=0cm 0cm 0cm 0cm, clip=true, width=0.3\linewidth]{
		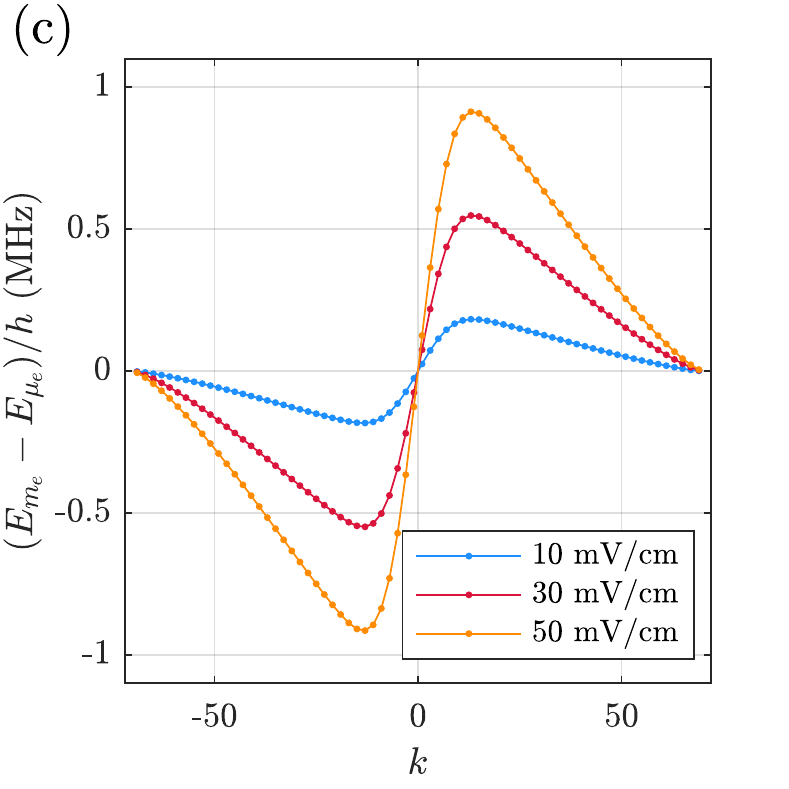}}
	\caption{
	(a) Fractional error in the dipole matrix elements at $n=(30,50,70)$ made when using the electron mass $m_e$ instead of the reduced mass $\mu_e$ for H$_2$ (circles), He$_2$ (triangles) and Na (stars) as a function of $\ell$. Only errors larger than 0.001 are shown. (b) Distribution of the $\ell$ character in the $(n=70,M_N=0)$ manifold of H$_2$, assuming zero quantum defect for all states. The character of the highest-$\ell$ states is located in the middle of the Stark manifold. (c) Resulting error from using $m_e$ instead of $\mu_e$ on the position of the Stark states in the $(n=70,M_N=0)$ manifold of H$_2$ for different fields, again assuming zero quantum defect for all states.	\label{fig:mu}
	}
\end{figure*}

\subsection{Calculation of Rydberg-Electron Wavefunctions and Radial Matrix Elements}
\label{sec:numerov}
\vspace*{5pt}

The radial integrals $\braket{n\ell | r | n'\ell'}$ were calculated numerically, following the general procedure described in Ref.~\cite{vsibalic17a}. For each wavefunction $\ket{n\ell}$, the radial Schr\"odinger equation of the Rydberg electron was solved using Numerov's method. The integration of the radial Schr\"odinger equation was performed in a pure Coulomb potential for each state at the corresponding zero-field energy 
\begin{align}
E = -\frac{hc\,\mathcal{R}_{\mathrm{H}_2}}{(n-\mu)^2} \,,
\end{align}
obtained as described in Section~\ref{sec:Hzero}, where $\mu$ is the quantum defect. The wavefunction was propagated from large $r$ values inwards towards the ion core until either the polarisability radius $r_\alpha=3.1664\,a_0$ of the H$_2^+$ ion \cite{jacobson97a} was reached or, for higher-$\ell$ states, until the wavefunction started to diverge beyond the inner classical turning point. For the integration, we used a square-root (sqrt) scaling of the radial coordinate \cite{bhatti81a} and all integrations were performed on the same grid. Consequently, no interpolation was necessary to compute the radial integrals.

The numerical error from the Numerov method was estimated by comparing radial matrix elements $\braket{n,\ell | r | n,\ell-1}$ computed with the same algorithm for the hydrogen atom to their analytical value $3n \sqrt{n^2-\ell^2}/2 $. We found fractional errors smaller than $10^{-9}$ and $10^{-10}$ for sqrt step sizes of 0.01~$a_0^{1/2}$ and 0.005~$a_0^{1/2}$, respectively. The corresponding shifts of the positions of the Stark states is of the order of several Hz and smaller step sizes do not decrease the fractional error further. We note that using a logarithmic scaling, as done by Zimmerman et al. \cite{zimmerman79a}, required doubling the number of grid points to achieve the same precision as with the sqrt scaling.

\subsection{Influence of the Reduced Mass}
\vspace*{5pt}

We use the reduced mass $\mu_e$ of the electron-core system in the Schr\"odinger equation, as done explicitly in Ref.~\cite{vsibalic17a}, which is essential when calculating Stark maps in light systems at high precision. Fig.~\ref{fig:mu}a shows the fractional error in the dipole matrix element $d=\braket{n,\ell |r| n,\ell-1}$ at $n=30$, 50 and 70 resulting from using the electron mass in the Numerov integration ($d_{m_e}$) compared to the result obtained with the reduced mass ($d_{\mu_e}$). For each calculated wavefunction, the error made in the inward integration is largest beyond the classical inner turning point, where the convergence behaviour is very sensitive to the energy at which the integration is performed. For increasing values of $\ell$, the inner turning point lies at larger $r$ values and therefore the error in the integral of the dipole matrix element rapidly increases. The errors are shown in Fig.~\ref{fig:mu}a as soon as they surpass 0.1\% and are displayed for the three systems of masses {2\,u} (H$_2$), {8\,u} (He$_2$) and {21\,u} (Na). In the case $n=70$ and $m=2\,$u, the error in the values of $d$ reaches 8\% for $\ell=69$. 

Fig.~\ref{fig:mu}b displays the distribution of the $\ell$ character in the hydrogenic $(n=70,m=0)$ Stark manifold at an arbitrary field. $\ell$ is not a good quantum number and the states are labelled with the integer number $k=n_1-n_2$, where $n_1$ and $n_2$ are the parabolic quantum numbers that arise in the solution of the Schr\"odinger equation in parabolic coordinates \cite{bethe57a}. The $\ell$ character is obtained from the frame transformation $\braket{n\ell m | nkm}=\braket{\ell | k}$, for which the Legendre polynomials represent a good approximation in the case of the lowest values of $\ell$ \cite{gallagher94a}. This approximation breaks down for higher values of $\ell$ and the high-$\ell$ character is increasingly localised in the middle of the Stark manifold. One can therefore expect the states in the middle of the Stark manifold to be more sensitive to the error made when approximating the reduced mass by the electron mass. The influence of this approximation on the positions of the $n=70$ Stark states of H$_2$ at different fields is depicted in Fig.~\ref{fig:mu}c, which was obtained after setting all quantum defects to zero for clarity. High-field- and low-field-seeking states are shifted to lower and higher energies, respectively. The absolute error increases towards the centre of the manifold, where it abruptly decreases again, because the states in the centre are only weakly affected by the field. In conclusion, failing to include the reduced mass in the radial Schr\"odinger equation leads to errors in the position of Stark-shifted high-$\ell$ states of H$_2$. These errors, though small, are significant at the applied fields (10-50~mV/cm) and experimental precision (better than 50~kHz) of the present study.

\section{Computational and Experimental Results}
\label{sec:results}

\subsection{Calculated Stark Maps}
\label{sec:maps}

\begin{figure}
	\centering
	\subfigure{\includegraphics[trim=0cm 0cm 0.5cm 0cm, clip=true, width=\linewidth]{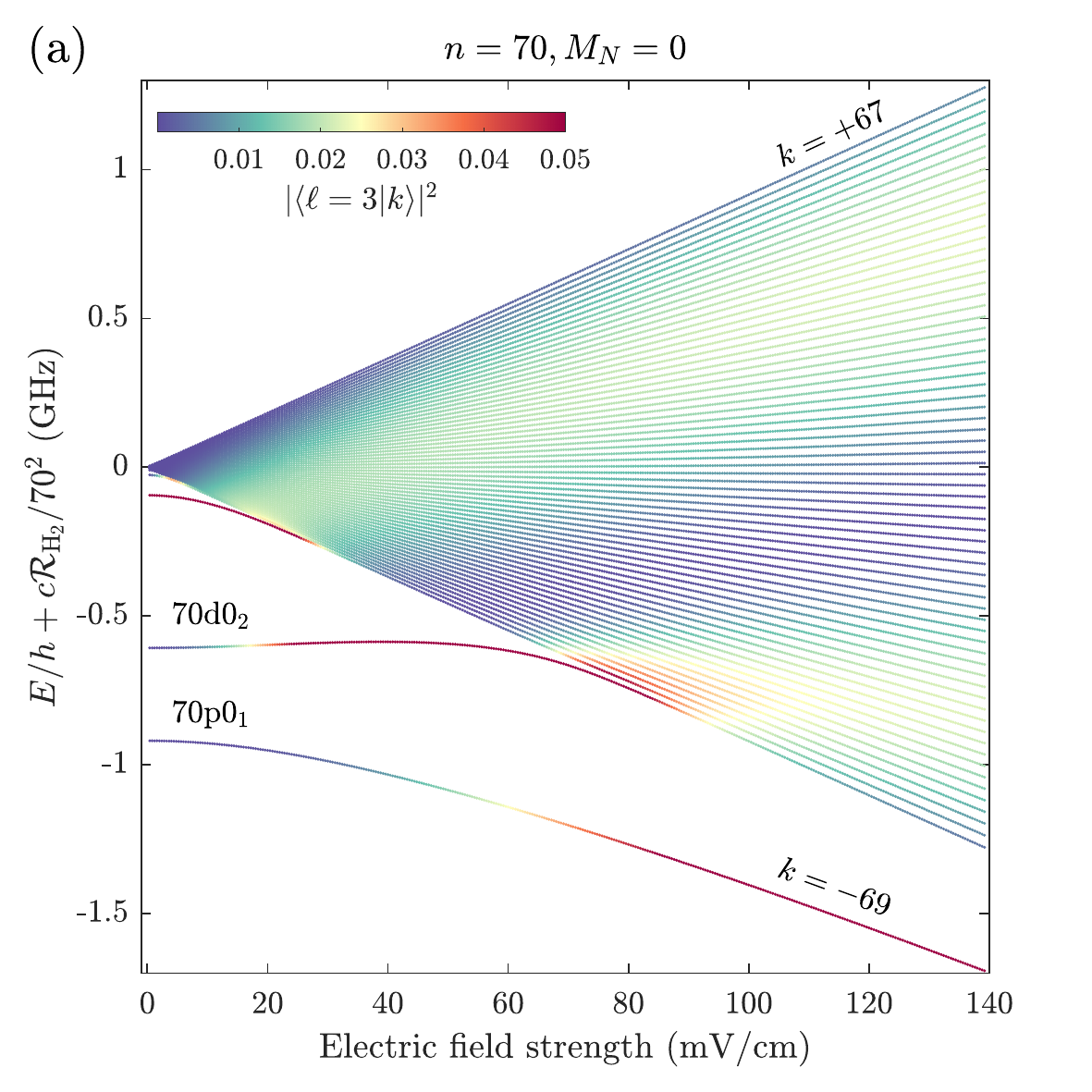}}
	\subfigure{\includegraphics[trim=0cm 0cm 0.5cm 0cm, clip=true, width=\linewidth]{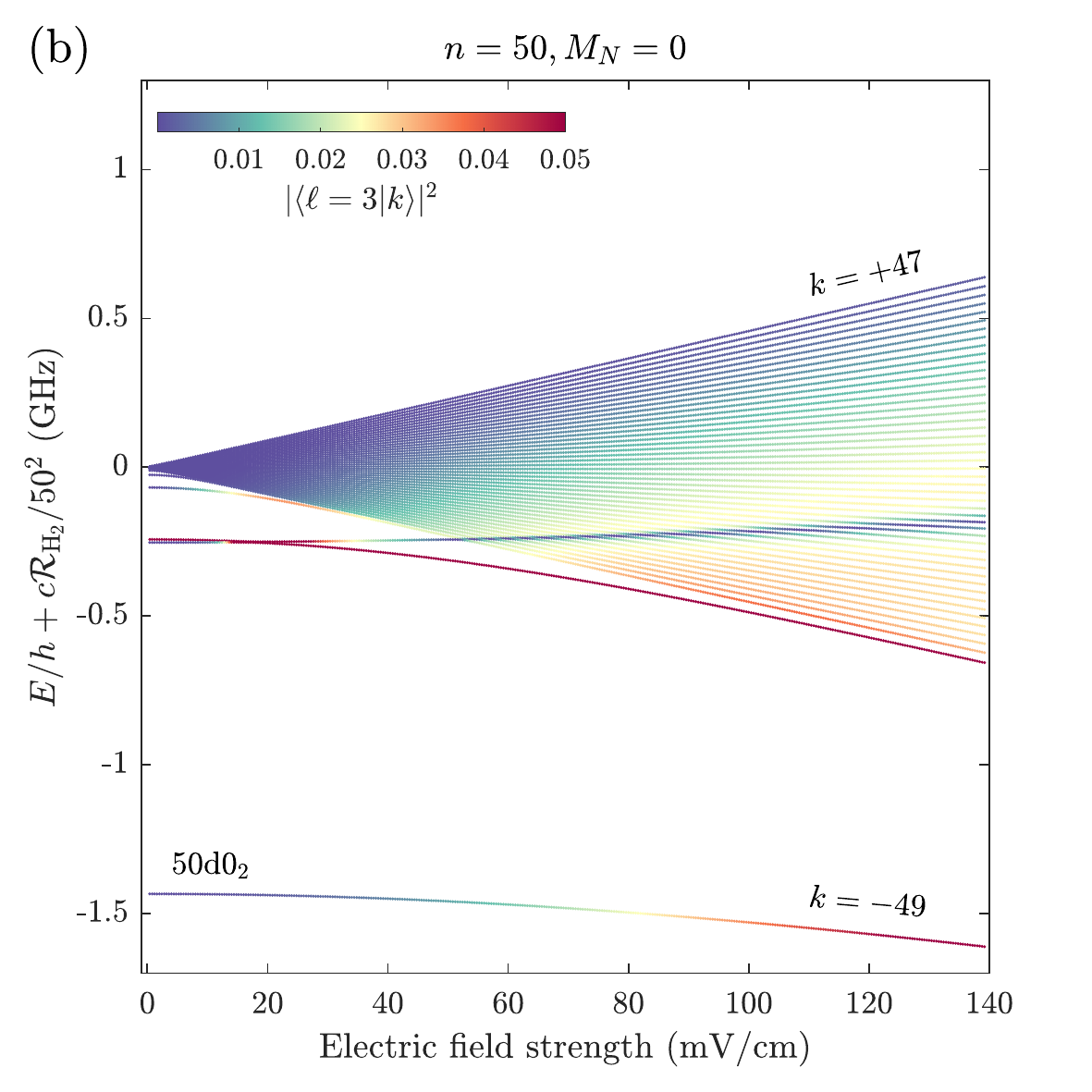}}
	\caption{Calculated Stark maps for (a) $n=70,M_N=0$ and (b) $n=50,M_N=0$ below the first ionisation threshold of para-H$_2$, where the position of zero quantum defect is placed at the origin of the frequency scale. The colour scale, which was capped at 0.05, indicates the amount of $n\mathrm{f}0_3$ character of each $k$ state in the field. The states are labelled using $k$ in the high-field limit, as indicated for the highest and lowest states depicted. The $n\mathrm{s}0_0$ states are located above the manifolds and are not visible in the figure. The lowest three states are, in order of increasing frequency at zero field: ($70\mathrm{p}0_1$, $70\mathrm{d}0_2$, $70\mathrm{f}0_3$) and ($50\mathrm{d}0_2$, $50\mathrm{p}0_1$, $50\mathrm{f}0_3$), respectively. For $n=50$, the p and f state form an avoided crossing at around 18~mV/cm.}
	\label{fig:colormanifold}
\end{figure}

\begin{figure*}[ht]
	\centering
	\subfigure{\includegraphics[trim=0.5cm 1.0cm 0.8cm 0.5cm, clip=true, width=0.47\textwidth]{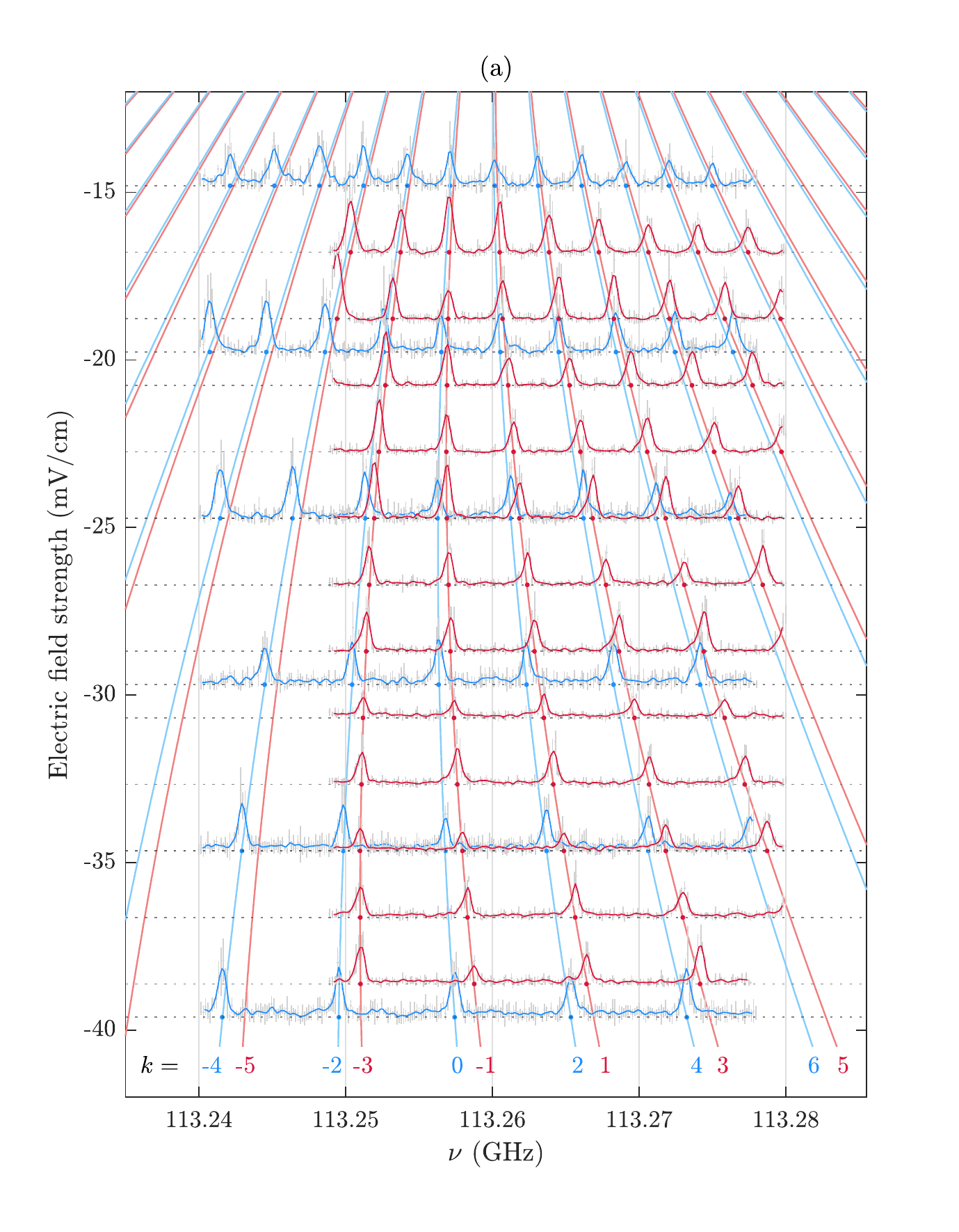}}
	\subfigure{\includegraphics[trim=1.2cm 1.0cm 0.1cm 0.5cm, clip=true, width=0.47\textwidth]{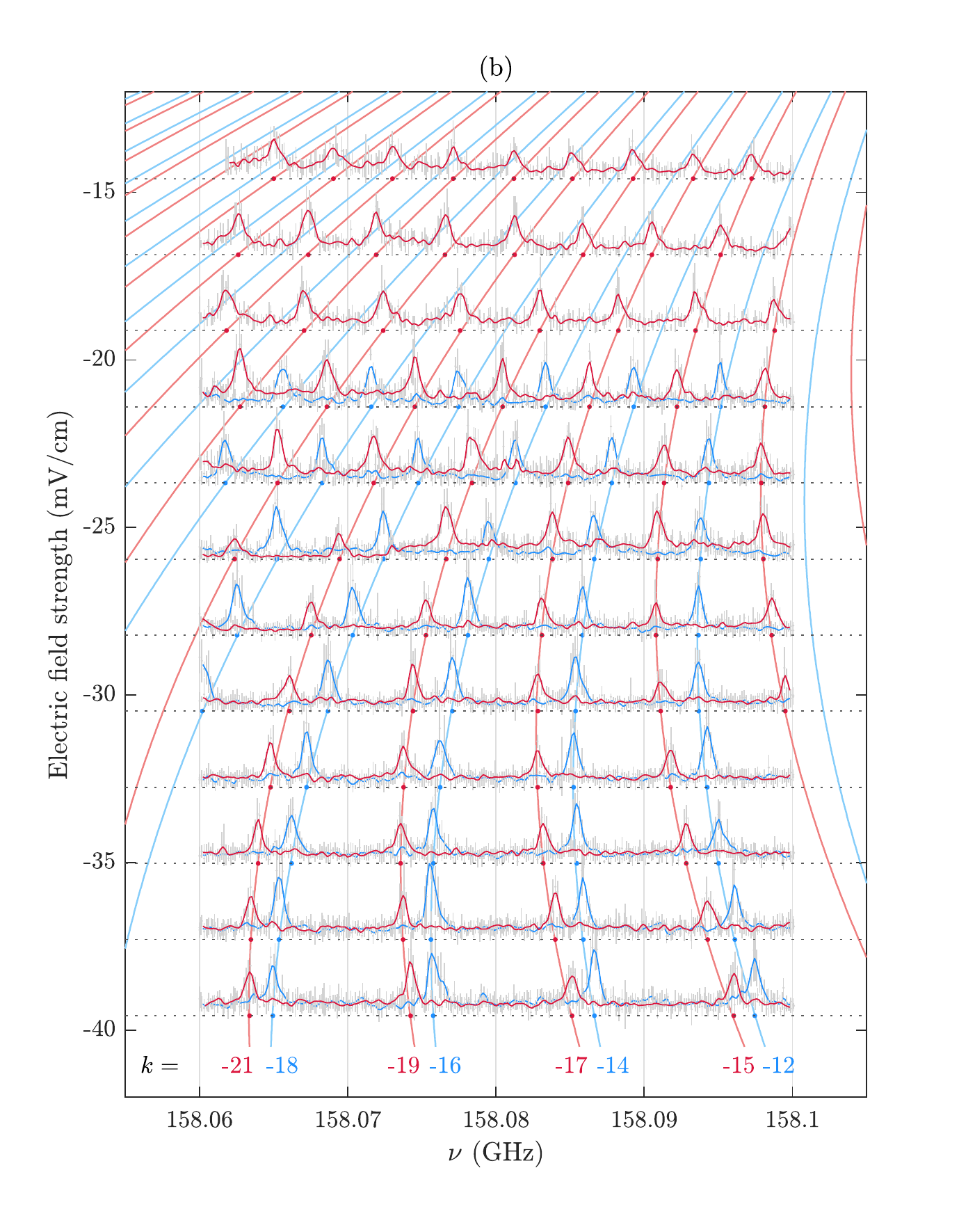}}
	\caption{
		Spectra of millimetre-wave transitions to the (a) $n=50$ and (b) $n=70$ Stark manifolds of H$_2$ with an X$^+\,^2\Sigma^+_g(N^+=0,v^+=0)$ core, recorded from the initial states $48\mathrm{d}0_2$ and $63\mathrm{d}0_2$, respectively, at various fields. The spectra were recorded with parallel polarisation of the mmW radiation. The polarisation of the UV laser exciting the initial d states was adjusted to populate either $M_N=0$ (red) or $M_N=1$ (blue) substates. The averaged raw data with vertical error bars are drawn in grey and the coloured spectra represent a running average of three data points. The corresponding field strengths are indicated by horizontal dashed lines and the coloured points indicate the fitted line centres. The calculated transition frequencies, obtained from the Stark maps of the initial and final states, are drawn in the corresponding colour, with the quantum number $k$ indicated for each state at the bottom.
		\label{fig:manifold-spectra}
	}
\end{figure*}

\fref{fig:colormanifold} shows the calculated Stark maps at (a) $n=70$ and (b) $n=50$ on the same field and frequency scale.  In both cases, the origin of the frequency scale was set at $-c\mathcal{R}_{\mathrm{H}_2}/n^2$, which we will from now on refer to as the \emph{zero-quantum-defect position}. 

In the experiment, the Stark spectra were recorded from $n$d$0_2$ levels and, consequently, the f character of the Stark states determines their intensities to a good approximation because the p $\leftarrow$ d Rydberg-Rydberg transitions are much weaker. Moreover, the p states are subject to predissociation.
The colour scale indicates the amount of $n\mathrm{f}0_3$ character of each state, obtained from the eigenvector matrix with elements $\braket{nkm | n\ell N^+_N}$ of the Stark Hamiltonian. It is capped at 0.05 in order to show the distribution of the character over the states of the manifold. In both cases, the $n\mathrm{s}0_0$ state is located above the manifold outside the depicted energy ranges. In the rest of this work, the Stark-shifted states are labelled in the high-field limit using the label $k$. In \fref{fig:colormanifold}b, for instance, the values of $k$ range from +47 for the highest to $-49$ for the lowest depicted state in steps of two. The 50s$0_0$ level corresponds to the $k=+49$ Stark state.

In the case of $n=70$, all states with $\ell>0$ exhibit decreasing quantum defects with increasing value of $\ell$. Because the $70\mathrm{d}0_2$ state lies closer to the p state than to the f state, it is first blue-shifted before being red-shifted and merging into the manifold as the field increases. At low fields, the f state determines the intensity distribution. As the field increases, the f character is first transferred to the high-$\ell$ manifold and to the d state, which starts dominating the intensity distribution at around 30~mV/cm. Beyond 90~mV/cm, it is the nominal p state ($k=-69$) that gains the largest f character. At low fields, the Stark states with the smallest f character are the most blue-shifted Stark states as well as the states that form a broad stripe in the middle of the manifold where the d character is concentrated.

In the case of $n=50$, the strong variation of the $n$p$0_1$ quantum defects caused by rotational channel interactions (see \fref{fig:lufano}) results in the p state lying only 10~MHz below the f state. This leads to an avoided crossing predicted at around 18~mV/cm. The distribution of f character follows the same overall behaviour as seen for $n=70$, the main differences being that (i) the d state gains f character at higher fields, and (ii) the p state leads to a sharp region of low f character in the manifold that is seen as a dark stripe in \fref{fig:colormanifold}b.

For each calculated Stark map, states with principal quantum number from $n-2$ to $n+2$ were included in the basis for the matrix diagonalisation. This basis proved large enough to achieve convergence on the kHz scale for the spectral positions of the Stark states at the applied fields.

\subsection{Spectra of the Stark Manifold}
\vspace*{5pt}

mmW spectra of transitions to the $n=50$ and 70 Stark manifolds recorded from the 48d$0_2$ and 63d$0_2$ states, respectively, are presented in \fref{fig:manifold-spectra}. The polarisation of the UV laser light was set so as to either excite the $|M_N=0|$ (parallel, in red) or $|M_N=1|$ (perpendicular, in blue) substates of the corresponding $n$d$0_2$ state. Each spectrum was normalized and is displayed with its baseline shifted vertically to the corresponding field value, indicated by horizontal dotted lines. The raw data are depicted in grey and the coloured lines represent running averages over three data points. A series of Lorentzian profiles was fitted to the spectrum and the fitted positions are indicated on the corresponding baselines as coloured dots. The centre positions were determined with accuracies of about 30~kHz and thus the error bars are too small to be visible at the scale of \fref{fig:manifold-spectra}. The coloured lines in vertical direction correspond to the calculated transition frequencies obtained from the Stark maps shown in Sec.~\ref{sec:maps} and include the quadratic shifts of the initial $n$d$0_2$ states which are superimposed on the linear shifts of the Stark manifolds. The calculated and experimental positions are in excellent agreement as quantified in Sec.~\ref{sec:fit}. Several parameters had to be introduced and optimised in order to reach a satisfactory agreement between experiment and calculations: (1) the quantum defects and Stark shifts of the initial n$d$ states, (2) a scaling parameter for the applied field of the order of 0.99, accounting for possible deviations of the effectively applied fields from the nominal ones, and (3) an uncompensated stray field in $z$ direction of less than 0.2~mV/cm.

The greatest challenge in resolving these transitions was the need to attenuate the mmW radiation to avoid power broadening of the lines. Significantly more power is needed to populate the components of the Stark manifold (attenuation of less than 20~dB) than to populate the $n$f states under field-free conditions (attenuation of more than 45~dB). The transition moments from the selected $n$d$0_2$ states to the Stark manifolds change rapidly with the value of $k$ because (i) the $n$f character varies significantly across the manifold (see \fref{fig:colormanifold} and discussion in the previous section) and (ii) the initial d states are polarised by the applied fields, which enhances the variation of the transition moment with $k$. 
It was therefore necessary to regularly adjust the mmW power between individual lines while recording the spectra. Consequently, the intensities depicted in \fref{fig:manifold-spectra} do not reflect the actual transition moments. 

\section{Sensitivities of the Stark-State Positions}
\label{sec:sensitivities}

\begin{figure}[ht]
    \centering
    \includegraphics[trim=0.5cm 1cm 0.5cm 0.5cm, clip=true, width=\linewidth]{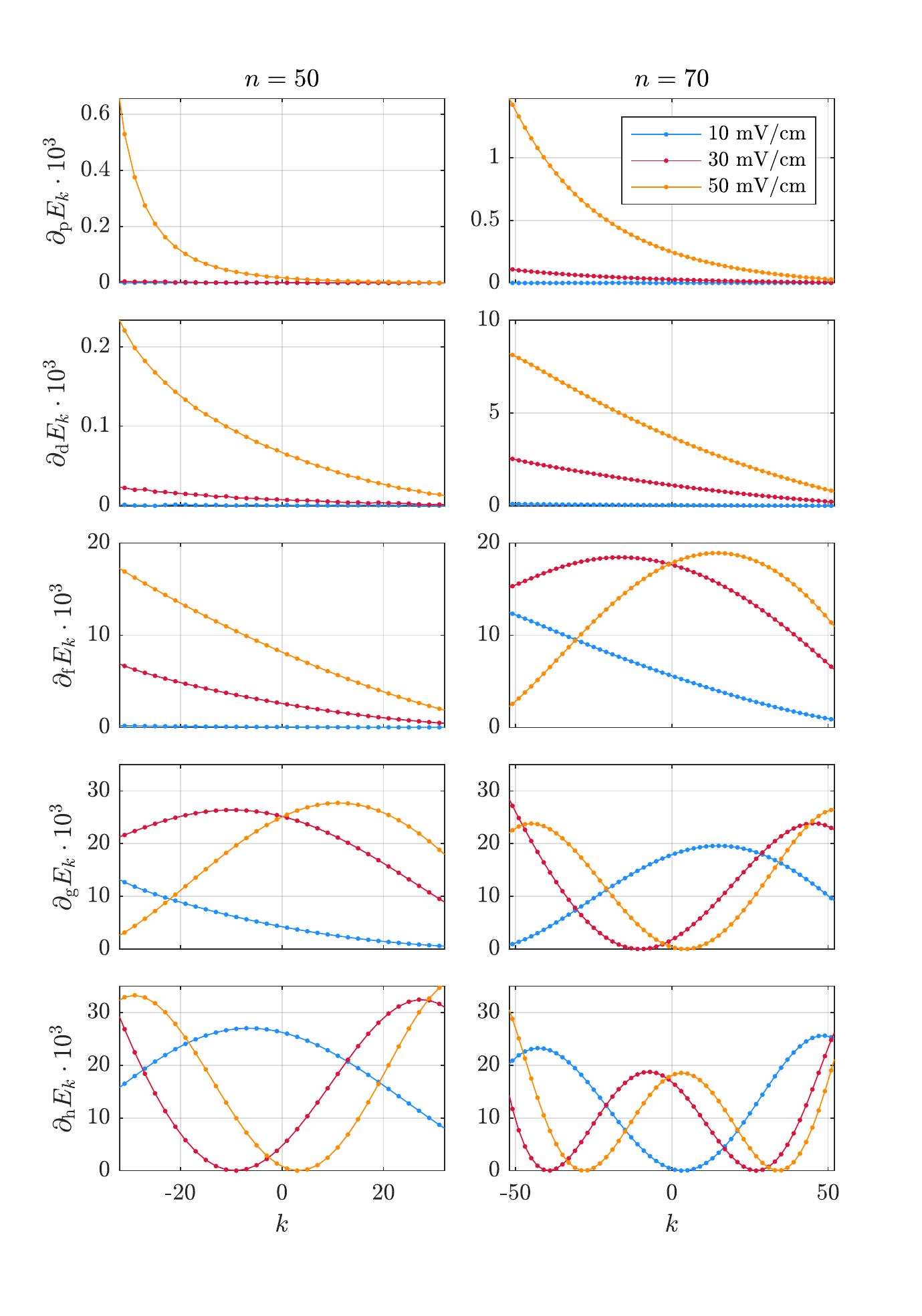}
    \caption{Sensitivities of the Stark states of H$_2$ in the linear manifold to a change of 1~MHz of the zero-field energy of different states with $1\leq\ell\leq5$ for $n=50$ (left column) and $n=70$ (right column), both with an X$^+\,^2\Sigma^+_g(N^+=0,v^+=0)$ core.}
    \label{fig:sensitivity}
\end{figure}

As stated in the introduction, the states of the Rydberg Stark manifold are desirable spectroscopic targets because their positions $E_k(F)$ are only weakly sensitive to uncertainties in the positions $E^0_\ell$ of the zero-field states. In order to quantify this statement, we varied the zero-field positions of the states with $1\leq\ell\leq5$ individually by 1~MHz and determined the resulting changes in the computed positions of the Stark states. In this way, we obtained energy sensitivities $\partial_\ell E_k(F)=\partial E_k(F) / \partial E^0_\ell$ at a given field strength $F$. The results are shown in \fref{fig:sensitivity} for $F=10$, 30 and 50~mV/cm, corresponding to the range of fields applied experimentally. In \fref{fig:sensitivity}, the left panels depict the central $n=50$ Stark states with $k\in\left[-30,30\right]$ and the right panels depict the central $n=70$ Stark states with $k\in\left[-50,50\right]$. As expected from perturbation theory, the further a state is from the manifold, the smaller is its influence on the manifold positions. Consequently, we can safely neglect uncertainties in the positions of the $n$s$0_0$ states, because they lie far above the manifold ($\mu\approx -0.1$) and are hardly mixed with manifold states at the fields of interest here. 

Using \fref{fig:colormanifold} as a reference, we see that the sensitivities are directly correlated to the corresponding $\ell$ character of the states. For instance at $n=70$ and a field of 50~mV/cm, the Stark states that are most sensitive to the zero-field position of the f state are the states with $k$ between 0 and 30 (see orange points in middle panel in the right column of  \fref{fig:sensitivity}). These states correspond to a region of high f character in \fref{fig:colormanifold}a. As soon as a low-$\ell$ state is merged into the manifold, its character is distributed over many $k$ states and the maximal sensitivities stop changing with the field strength. For instance, the positions of the Stark states at $n=70$ are less sensitive to $E^0_\mathrm{f}$ at 10~mV/cm, where the f state is not yet fully immersed in the manifold, than at fields above 20~mV/cm. Beyond 20~mV/cm, the maximal sensitivity reaches 0.02, i.e., 20~kHz for a change $\delta E^0_\mathrm{f}$ of 1~MHz.

In order to improve the ionisation energy of H$_2$ by one order of magnitude, we aim to reach a description of the manifold states with an uncertainty lower than 50~kHz. The sensitivities shown in \fref{fig:sensitivity} determine the accuracy needed in the zero-field energies, and thus in the quantum defects, of the low-$\ell$ states to reach this goal. We can determine the uncertainties in the quantum defects $\delta\mu$ corresponding to an uncertainty $\delta\epsilon / (hc)$ of 50~kHz using
\begin{align}
\delta\epsilon / (hc) \approx \frac{2\mathcal{R}_{\mathrm{H}_2}}{n^3}\delta\mu\,.
\end{align}
The resulting values and respective fractional uncertainties $\delta\mu/\mu$ are listed in Table~\ref{tab:qd}. At 50~mV/cm, states with $\ell\geq4$ for $n=50$ and $\ell\geq3$ for $n=70$ are merged in the manifold and therefore $\delta\mu$ becomes independent of $\ell$.

\begin{table}[h]
\centering
\caption{Error $\delta\mu$ and fractional error $\delta\mu/\mu$ in the quantum defect $\mu$ which would lead to shifts of manifold positions up to 50~kHz at fields up to 50~mV/cm (see also Figs.~\ref{fig:lufano} and \ref{fig:sensitivity}). \label{tab:qd}}
\begin{tabular}{c cc cc}
\\[-1.5ex]
\hline\hline
	& \multicolumn{2}{c}{$n=50$} &\multicolumn{2}{c}{$n=70$} \\
	\cmidrule(lr){2-3} \cmidrule(lr){4-5}
	& $\delta\mu$ & $\delta\mu/\mu$ & $\delta\mu$ & $\delta\mu/\mu$ \\
	\hline 
	\\[-1.5ex]
	$n$p$0_1$ &$9.5\times10^{-3}$  & 1.98 & $1.0\times10^{-3}$ & 0.02 \\
	$n$d$0_2$ & $9.5\times10^{-3}$ & 0.35 & $5.2\times10^{-4}$ & 0.02 \\
	$n$f$0_3$ & $9.5\times10^{-5}$  & 0.02 &  $1.0\times10^{-4}$ & 0.02 \\
	\\[-2ex]
	$n$g$0_4$ & $3.8\times10^{-5}$ & 0.03 & $1.0\times10^{-4}$ & 0.08 \\
	$n$h$0_5$ & $3.8\times10^{-5}$ & 0.08 & $1.0\times10^{-4}$ & 0.21 \\
	$n$i$0_6$  & $3.8\times10^{-5}$ & 0.18 & $1.0\times10^{-4}$ & 0.49 \\
\hline\hline
\end{tabular}
\end{table}

In previous work \cite{osterwalder04a, beyer19a, sprecher14b}, the uncertainty $\delta\mu$ of the $n$p and $n$f quantum defects was estimated to be $1.6\times10^{-5}$, corresponding to an uncertainty of 500~kHz at $n=60$. However, the $70$p$0_1$ state lies close to a vibrational perturber (see Sec.~\ref{sec:Hzero}) and we estimate that the uncertainty in its quantum defect is larger than for other members of the $n$p$0_1$ series. Moreover, comparisons of our measured spectra to calculated Stark shifts of initial and final states revealed that the $n$d$0_2$ positions predicted by the MQDT calculations \cite{sprecherDiss} must be modified significantly in order to reach an acceptable agreement with the experimental data, e.g., by almost 60~MHz at $n=48$. This corresponds to an uncertainty in the quantum defects of $1\times10^{-3}$, which would limit the accuracy of the calculated Stark states at $n=70$ to about 100~kHz (see \tref{tab:qd}). To overcome these limitations, we determined the positions of the $70$p$0_1$ and $70$d$0_2$ states relative to the  $70$f$0_3$ state with a precision of better than 1~MHz by laser spectroscopy \cite{doran22a}.

We estimate our quantum defects for states with $\ell>4$ and $n=50,70$ to be accurate to better than 10\%, as was discussed in Sec.~\ref{sec:Hzero}. Referring to \tref{tab:qd}, one also sees that for $\ell\geq5$ uncertainties in the quantum defects do not limit the accuracy of the calculated Stark-state positions for $n\geq50$. Consequently, the quantum defects of the g states are the most critical, especially at $n=50$, and we return to this point in the next section when we discuss the systematic uncertainties of our final results.

\section{Analysis of Measured Spectra}

\subsection{Quantitative Comparison with Experiments}
\label{sec:fit}

\begin{figure}
	\centering
	\subfigure{\includegraphics[trim=0cm 1.5cm 1cm 1cm, clip=true, width=\linewidth]{
		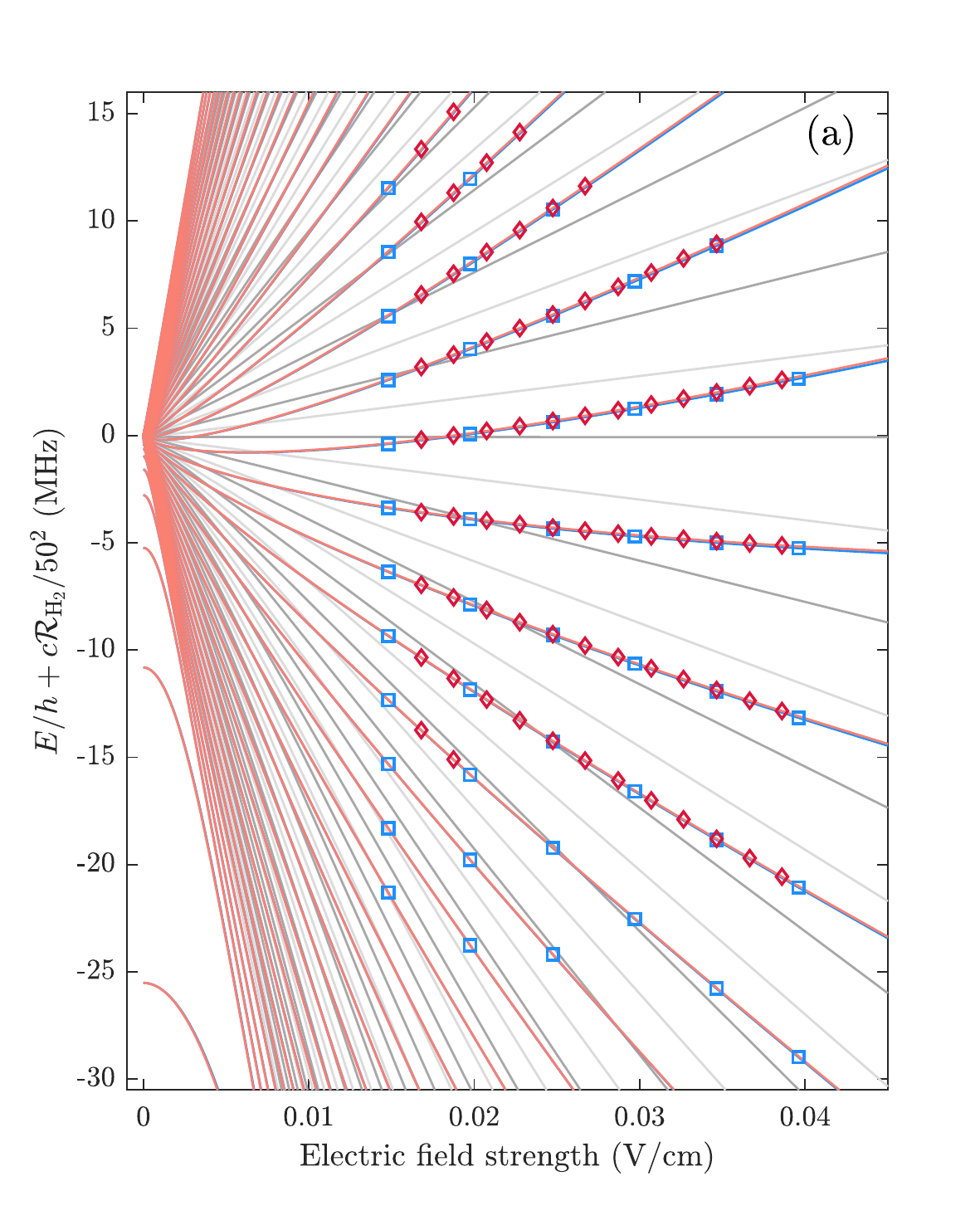}}
	\subfigure{\includegraphics[trim=0cm 0.88cm 1cm 0.19cm, clip=true, width=\linewidth]{
		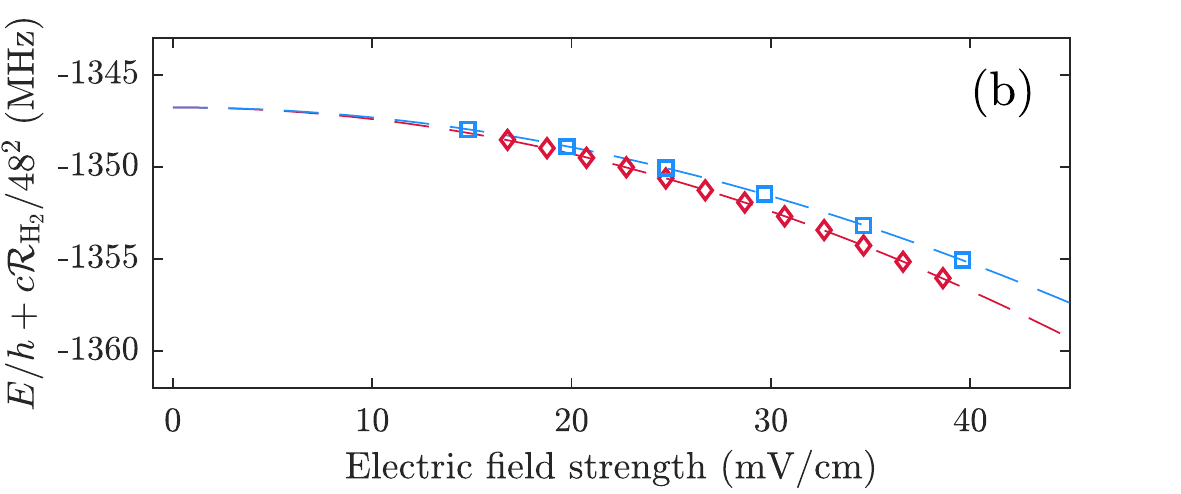}}
	\subfigure{\includegraphics[trim=0cm 0cm 1cm 0.2cm, clip=true, width=\linewidth]{
		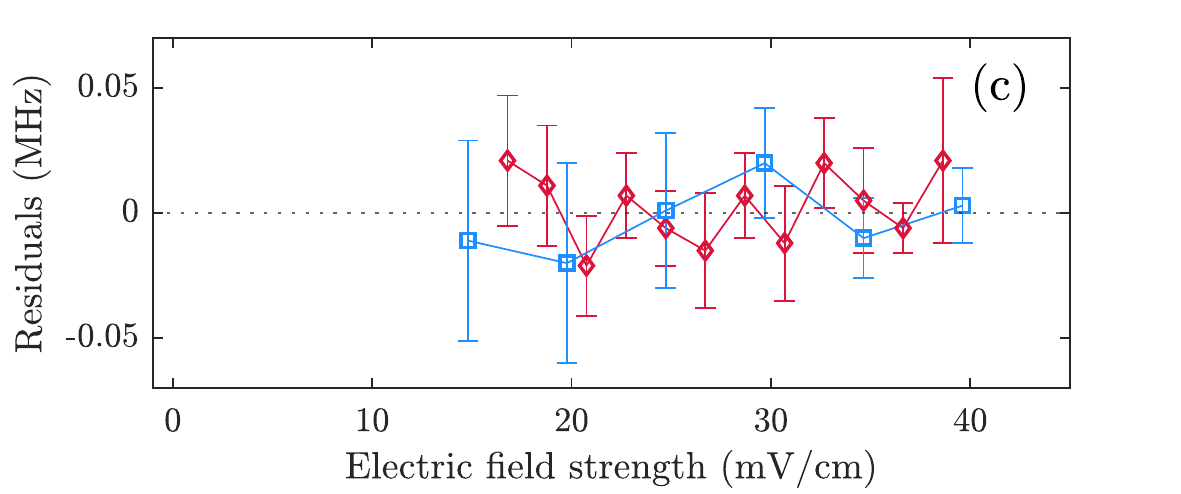}}
	\caption{(a) Calculated Stark map for $n=50$ in the region of the measured $k$ states with the origin of the energy axis set at the position of zero quantum defect. The states with $M_N=0$ and $M_N=1$ are drawn as red and blue lines, respectively. They are almost degenerate and therefore only distinguishable at the highest fields. For comparison the near-hydrogenic Stark states with $M_N=10$ and $M_N=11$ are drawn in dark and light grey, respectively. The state with a field-free position of $\approx -25\,$MHz at the chosen scale is 50h$0_5$. Red diamonds and blue squares mark at which fields and to which $k$ states transitions were recorded (see \fref{fig:manifold-spectra}). (b) Averaged shifts of the initial states 48d$0_2(M_N=0,1)$ (open squares and diamonds) obtained by subtracting the transition frequencies from the calculated positions of the Stark manifold. Fitted Stark shifts are drawn as dashed lines, obtained by simultaneously fitting two polynomials with the same field-free position. (c) Corresponding residuals of the combined fit (see text for details).}
	\label{fig:dshift-map}
\end{figure}

\fref{fig:dshift-map}a presents the calculated maps of the $M_N=0$ (red lines) and $M_N=1$ (blue lines) Stark manifolds for $n=50$ obtained with our best estimates of the quantum defects (see Supplementary Material). In the depicted field and energy ranges, the two manifolds are almost degenerate and they cannot be distinguished on the scale of the figure, except at the highest fields. In hydrogenic Rydberg systems, the Stark states with odd and even values of $M_N$ do not coincide but appear equally spaced and alternating between even and odd $M_N$ values, as depicted by the dark and light grey lines, which correspond to $M_N=10$ and $M_N=11$, respectively. These states are superpositions of $\ket{\ell M_N}$ states with $\ell\geq10$ and $\ell\geq11$, which all have quantum defects of less than $2\times10^{-5}$ (i.e., shifts of less than 1~MHz at $n=50$), and are thus almost hydrogenic. The reason for the near degeneracy of the $M_N=(0,1)$ manifolds is that the 50s$0_0$ state, which only contributes to the $M_N=0$ manifold, lies more than 5~GHz higher and does not mix significantly with other states at the investigated fields. The observed $k$ states in both manifolds are therefore almost identical superpositions of $\ell\geq1$ states. Unlike the near-hydrogenic $M_N=(10,11)$ manifolds, which exhibit an almost perfectly linear Stark effect, the $M_N=(0,1)$ manifolds show a pronounced upward curvature. This behaviour originates from the nonzero values of the quantum defects of states with $\ell<10$.

Given that the stray field and the scaling factor relating the nominal and the effectively applied fields were not known exactly, they had to be adjusted to best match the measured line spacings for each set $\left\{\nu_i(F,M_N)\right\}$ of transitions recorded at a given field strength $F$ and $M_N$ value. The index $i$ labels the final Stark states in the manifold which were accessed. We determined a value of 0.9925(8) for the field scaling factor, which is a property of the experimental setup, and values of 65(5)~$\mu$V/cm and 169(6)~$\mu$V/cm for the stray fields present in the photoexcitation volume on the days the $n=50$ and $n=70$ manifolds were recorded, respectively. The blue squares and red diamonds in \fref{fig:dshift-map}a designate the calculated positions of the Stark states for which the spectra shown in \fref{fig:manifold-spectra}a were recorded at the effectively applied fields. 

We subtract the measured transition frequencies (see \fref{fig:manifold-spectra}) from the calculated positions of the Stark manifold both for the {$M_N=0$} and {$M_N=1$} manifolds. Because these two manifolds were recorded from the distinct $M_N=(0,1)$ components of the 48d$0_2$ state, the subtraction procedure yields a single position for each set of transitions $\left\{\nu_i(F,M_N)\right\}$ with some scatter, which helps to assess the accuracy of the calculations. 

\begin{figure}
	\centering
	\subfigure{\includegraphics[trim=0.2cm 0.9cm 0.8cm 0cm, clip=true, width=\linewidth]{
		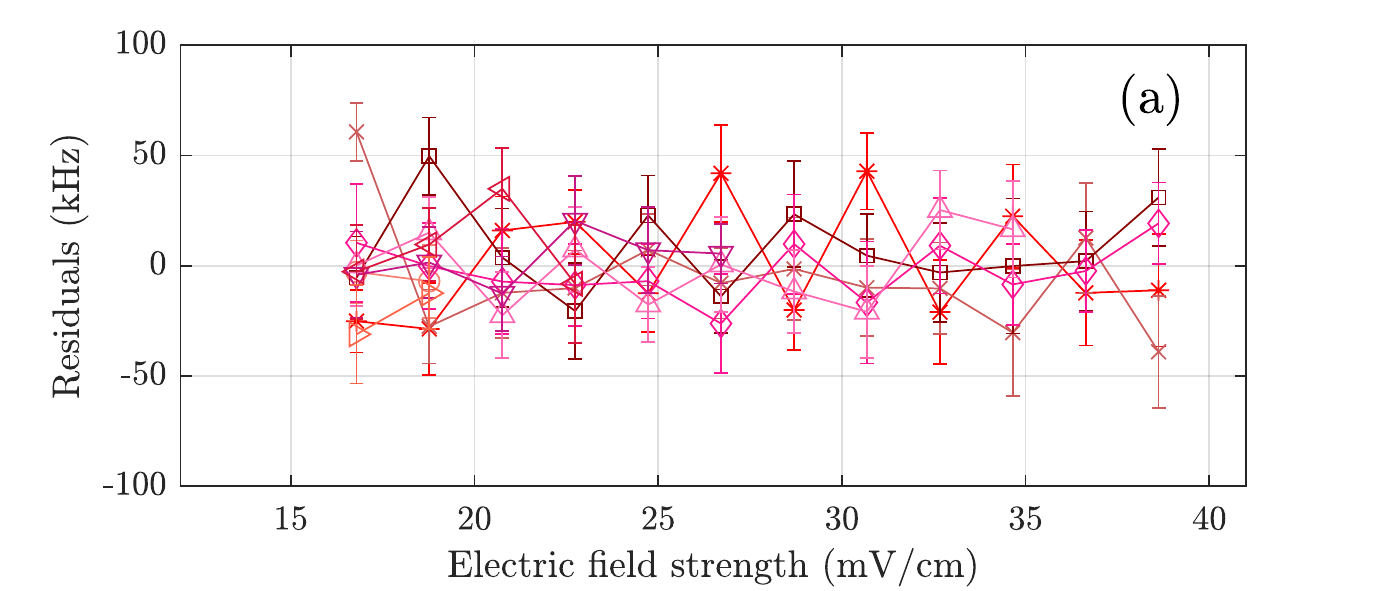}}
	\subfigure{\includegraphics[trim=0.2cm 0cm 0.8cm 0.4cm, clip=true, width=\linewidth]{
		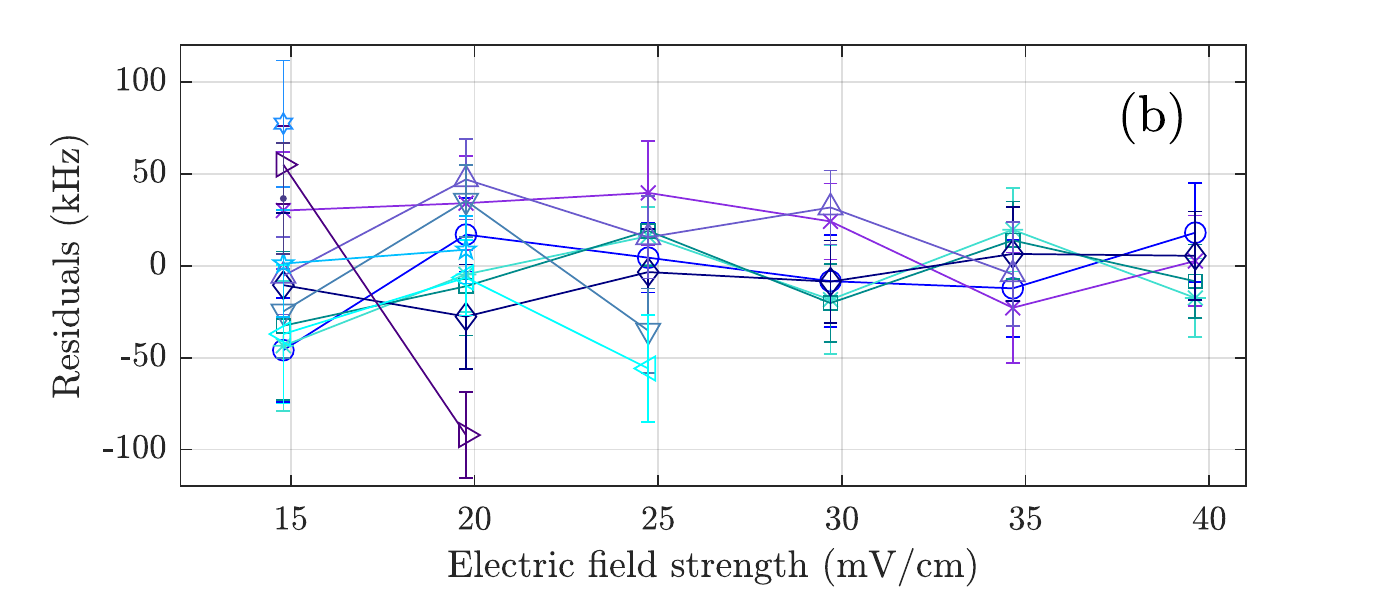}}
	\caption{Residuals of the determination of the Stark shift of the initial 48d$0_2$ state with (a) $M_N=0$ and (b) $M_N=1$. The residuals for each measured $k$ state are shown in different colours and symbols. At each field, they correspond to the deviation from the mean Stark shift of the initial state determined from all measured $k$ states.}
	\label{fig:dshift-residuals}
\end{figure}

\fref{fig:dshift-map}b depicts the field dependence of the initial 48d$0_2$ state determined for each field and $M_N$ value used in the experiment. The red diamonds $(M_N=0)$ and the blue squares $(M_N=1)$ represent averages over the values obtained from the different Stark states. The scatter of the data, characterised by the standard deviations, is much smaller than the size of the symbols. The residuals are represented on an enlarged scale in \fref{fig:dshift-residuals}, where the different colours encode the different $k$ states. The residuals are almost all smaller than 50~kHz and do not reveal systematic trends with field strength or $k$ value, neither for $M_N=0$ nor for $M_N=1$. We conclude that (i) the calculations describe the Stark shifts of the $n=50$ manifold with an accuracy of $\pm50$~kHz, and (ii) the experiments also provide accurate values of the Stark shifts of the 48d$0_2 (M_N=0,1)$ states. Similar conclusions were reached in the analysis of the $n=70$ Stark manifold, where the shifts of the 63d$0_2 (M_N=0,1)$ states were determined (see Supplementary Material for details). For each of the initial states $n^\prime$d$0_2$, the sublevels with different $M_N$ values are exactly degenerate at zero field and their energy is kept the same for both data sets. We perform weighted fits of polynomials
\begin{align}
\Delta\nu(F) = a_0 + a_2 \, F^2 + a_4 \, F^4
\label{eq:polynomial}
\end{align}
to the Stark shifts of all initial states, where $a_0$ is the $M_N$-independent field-free position relative to the zero-quantum-defect position of the measured manifold, and $a_2$ and $a_4$ are $M_N$-dependent polarizabilities. The red and blue lines in \fref{fig:dshift-map}b represent the Stark shift of the 48d$0_2$ states obtained with \eref{eq:polynomial} and the corresponding fit residuals are depicted in \fref{fig:dshift-map}c. \tref{tab:dshift} summarises the fit results obtained for all four measured manifolds, where we have converted $a_0$ into the binding energy of the $n^\prime$d$0_2$ state using the fact that the zero-quantum-defect positions of the manifolds are given by $-c\mathcal{R}_{\mathrm{H}_2}/n^2$. Individual fits of the Stark shifts at $n=50$ for both $M_N=(0,1)$ yield binding energies that agree within their combined uncertainty with the values in \tref{tab:dshift} for 48d$0_2$, which further supports the reliability of our analysis. In the case of 63d$0_2$, the individual fit of the $(M_N=1)$ dataset leads to a deviation of $+150$~kHz from the binding energy in \tref{tab:dshift}, which might indicate a larger influence of inhomogeneous or perpendicular stray fields at $n=70$, particularly at the lowest fields.

We remark that the direct calculation of the position of the Stark shifts of the initial $n^\prime$d$0_2$ states would not have reached the accuracy of the results summarised above because of the sensitivity of the d-state shifts at low fields on the relative positions of the adjacent p and f states. In this case, the field-free positions of the different p, d and f states are correlated parameters which makes accurate predictions difficult.

\begin{table*}[ht]
\centering
\caption{Results of the polynomial fit of Stark shifts of the initial $n^\prime$d$0_2(M_N)$ states (see \fref{fig:dshift-map}b). The parameters are the $M_N$-independent energies $\Delta E^0$ of the states $n^\prime$d$0_2$ relative to the corresponding zero-quantum-defect position and $M_N$-dependent polarisabilities. \label{tab:dshift}}
\begin{tabular}{c cccc}
\\[-1.5ex]
\hline\hline
	\\[-2ex]
	$n^\prime\ell N^+_N(M_N)$ &  $\Delta E^0/h$ (MHz) & $\Delta_\text{e-c}$ & $a_2$ (MHz mV$^{-2}$ cm$^2$) & $a_4$ (MHz mV$^{-4}$ cm$^4$)
	\\[0.1ex]
	\hline 
	\\[-1.5ex]
	$48\mathrm{d}0_2(0)$ & $-1346.762(21)$ & $+57.29$ & $-0.634(6)\times10^{-2}$ & $0.07(3)\times10^{-6}$ \\
	$48\mathrm{d}0_2(1)$ & $-1346.762(21)$ & $+57.29$ & $-0.544(5)\times10^{-2}$ & $0.10(3)\times10^{-6}$ \\
	\\[-2ex]
	$63\mathrm{d}0_2(0)$  & $-677.317(30)$  & $+18.07$ & $-4.084(8)\times10^{-2}$ & $1.09(4)\times10^{-6}$ \\
	$63\mathrm{d}0_2(1)$  & $-677.317(30)$  & $+18.07$ & $-3.431(7)\times10^{-2}$ & $0.71(4)\times10^{-6}$ \\
\hline\hline
\end{tabular}
\end{table*}

\begin{table*}[ht]
\centering
\caption{Field-free transition frequencies to $n$f$0_3$ and $n$g$0_4$ states from the corresponding initial $n^\prime$d$0_2$ states. With the values from \tref{tab:dshift}, their energies $\Delta E^0$ relative to the corresponding zero-quantum-defect position and quantum defects $\mu$ were determined. These values are compared to the values obtained from MQDT calculations and the differences are denoted as $\Delta_\text{e-c}$. \label{tab:bindingenergy}}
\begin{tabular}{c S[table-format=7.5] SSS[table-format=2.4, retain-explicit-plus] S[table-format=2.7]S[table-format=2.7]S[table-format=2.6, retain-explicit-plus]}
\\[-1.5ex]
\hline\hline
	\\[-2ex]
	& & \multicolumn{3}{c}{$\Delta E^0/h$ (MHz)} &\multicolumn{3}{c}{$\mu / 10^{-3}$} \\
	\cmidrule(lr){3-5} \cmidrule(lr){6-8}
	& \multicolumn{1}{c}{$\nu_\mathrm{exp}$ (MHz)} & \multicolumn{1}{c}{Experiment} & \multicolumn{1}{c}{Calculation} & \multicolumn{1}{c}{$\Delta_\text{e-c}$} & \multicolumn{1}{c}{Experiment} & \multicolumn{1}{c}{Calculation} & \multicolumn{1}{c}{$\Delta_\text{e-c}$}
	\\[0.1ex]
	\hline 
	\\[-1.5ex]
	$50\mathrm{f}0_3$ 	& 113021.32(3)	& -240.99(4) 	& -242.5(8) 	& +1.5(8) 	& 4.5789(8) 	& 4.608(16) 	& -0.029(16) \\
	$70\mathrm{f}0_3$ 	& 158028.89(8)	&  -93.59(8)  	&  -94.1(3)  	& +0.5(3) 	& 4.8797(21)  	&  4.905(16)  	&  -0.025(16)  \\
	\\[-2ex]
	$50\mathrm{g}0_4$ 	& 113192.9(5) 	&  -69.4(5)  	&  -70.5  		& +1.1(5)	& 1.319(10)	& 1.340 		& -0.021(10) \\
	$70\mathrm{g}0_4$ 	& 158096.4(5) 	&  -26.1(5)  	&  -25.7  		&  -0.4(5)  & 1.361(26)		& 1.345 		&  +0.016(26) \\
\hline\hline
\end{tabular}
\end{table*}

\subsection{Determination of the Quantum Defects of f and g states}
The binding energies of the 48d$0_2$ and 63d$0_2$ states, determined in the previous section, make it possible to derive the binding energies and quantum defects of all states that can be spectroscopically accessed from them. In this section, we focus on the $n$f$0_3$ and $n$g$0_4$ states at $n=50$ and $n=70$, in order to validate the positions we have used in our calculations. 

mmW spectra of the transitions to $n$f states can be recorded directly from the $n^\prime$d levels, as illustrated in Panels (c) and (d) of  \fref{fig:measurement}. We obtain the energies $\Delta E^0$ of the final states relative to the corresponding zero-quantum-defect position as
\begin{align}
\Delta E_{n\mathrm{f}}^0 = \Delta E_{n^\prime\mathrm{d}}^0 + h\nu_\mathrm{exp} - hc\mathcal{R}_{\mathrm{H}_2}\left(\frac{1}{n^{\prime 2}}-\frac{1}{n^2}\right)
\label{eq:balmer}
\end{align}
directly from the Balmer formula. \tref{tab:bindingenergy} compares the determined values for $\Delta E^0$ and the resulting quantum defects of the $n$f$0_3$ states with the values obtained from MQDT calculations with the quantum defects reported in Ref.~\cite{osterwalder04a}. The present results indicate that the error in the $n$f quantum defects was underestimated by a factor of two. 

\begin{figure}[ht]
	\centering
	\subfigure{\includegraphics[trim=0cm 0cm 0.5cm 0cm, clip=true,width=0.48\linewidth]{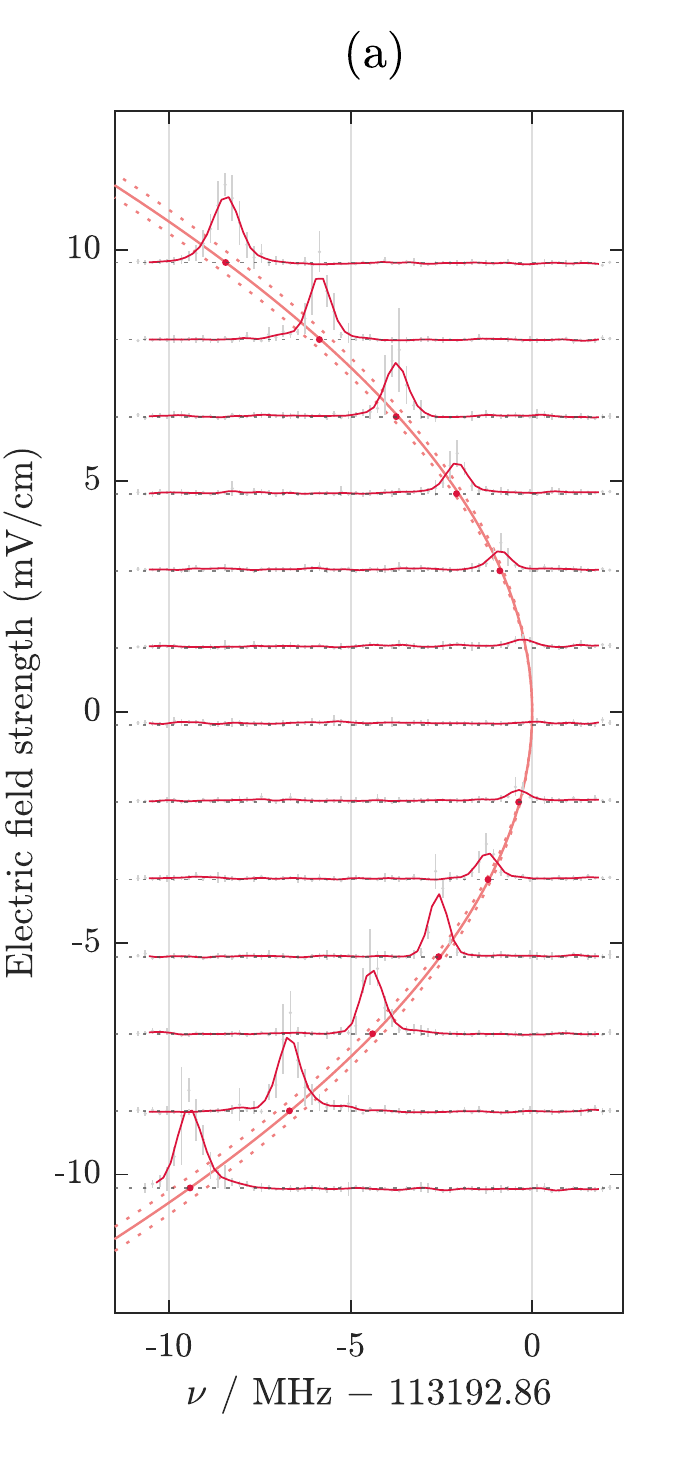}}
	\subfigure{\includegraphics[trim=0cm 0cm 0.5cm 0cm, clip=true,width=0.48\linewidth]{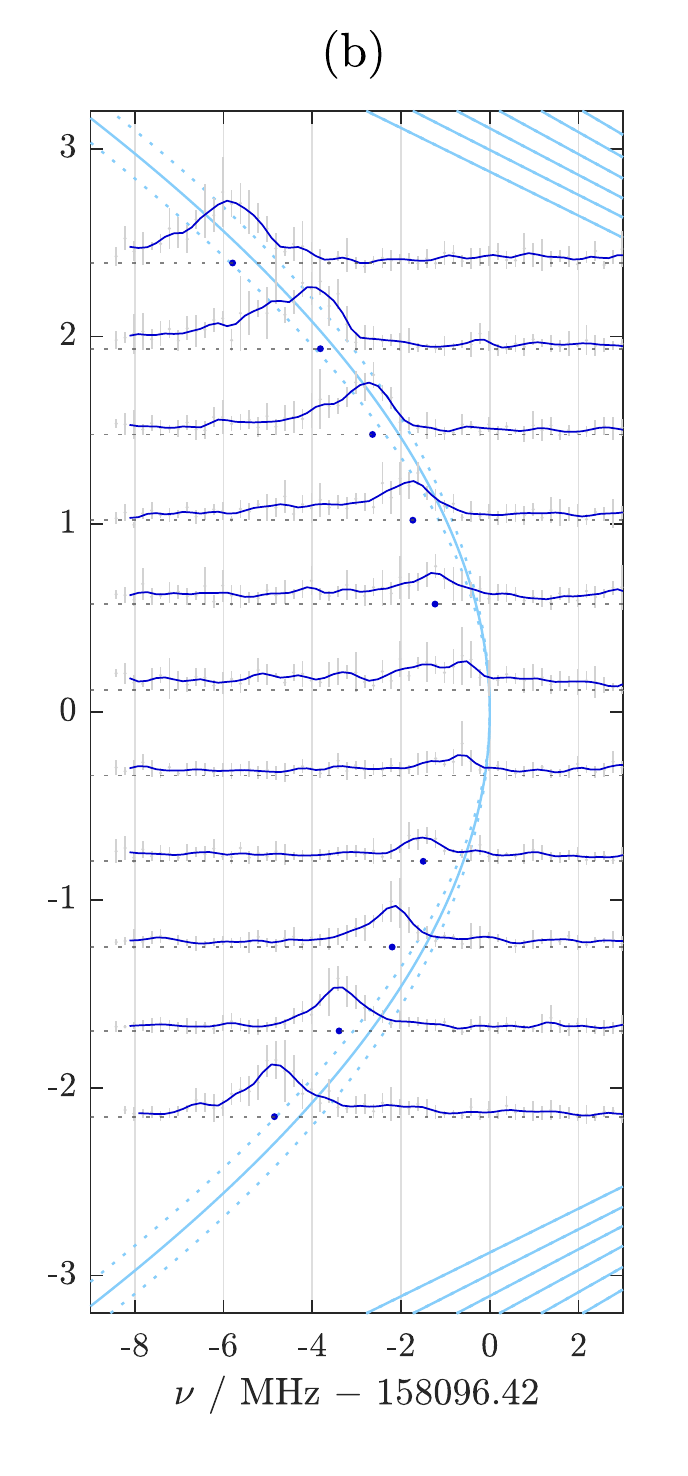}}
	\caption{Spectra of (a) the 50g$0_4\leftarrow$ 48d$0_2 (M_N=0)$ and (b) the 70g$0_4\leftarrow$ 63d$0_2 (M_N=1)$ transitions recorded in the presence of weak electric fields. The spectra were shifted vertically so that the origin of their intensity scale matches the value of the applied electric field given along the vertical axis and their fitted line centres are indicated by coloured dots. Light red and blue lines show the calculated Stark shifts with the field-free positions of the $n$g$0_4$ states determined from the spectra, to which the frequency scale is referenced. Dotted lines show the results if the corresponding $n$h$0_5$ state is shifted by $\pm2$~MHz.}
	\label{fig:quadratic}
\end{figure}

One-photon mmW transitions from the initial $n^\prime$d$0_2$ states to $n$g levels are dipole-forbidden under field-free conditions but become observable already at weak fields because the separation between the f and g states is small and the g states rapidly gain f character as the field increases. Panels (a) and (b) of \fref{fig:quadratic} display the Stark spectra around the field-free position of the transitions $50\mathrm{g}0_4(M_N=0)\leftarrow 48\mathrm{d}0_2(M_N=0)$ and $70\mathrm{g}0_4(M_N=1)\leftarrow 63\mathrm{d}0_2(M_N=1)$, respectively. The baseline of each spectrum is shifted vertically to the corresponding value of the applied field and line centres obtained from fitting asymmetric Lorentzians are indicated as dots. As expected, the line intensity increases with the magnitude of the Stark shift, which corresponds to a larger f character of the final state. The field-free transition frequencies to the g states were estimated from a quadratic fit of the line centres with an uncertainty of 500~kHz in both cases. The frequency scales in \fref{fig:quadratic} are referenced to the field-free positions of the respective transitions. In the case of 50g$0_4$, the intensity of the symmetric lines vanishes close to zero field, which is in contrast to the spectra for 70g$0_4$, where the lines are asymmetrically broadened and the intensities do not exactly vanish at the apex of the parabolic curve. We attribute both effects to inhomogeneous stray fields in the photoexcitation volume, which have a stronger impact at $n=70$ than at $n=50$ because of the larger polarisabilities and smaller fields applied. Analysis of the line shape in such situations \cite{osterwalder99a} indicates that the line centres are at the blue edges of the line profiles. Close to zero field, the observed intensity is exclusively field-induced.

Absolute binding energies of the $n$g$0_4$ states can be determined from these frequencies as for the $n$f$0_3$ states, see \eref{eq:balmer}, and the results are summarised in \tref{tab:bindingenergy}. The deviations from the calculated binding energies do not have an influence on the manifold positions at the 50\, kHz level.  In \fref{fig:quadratic}, light-red and light-blue lines indicate the Stark shifts of the transitions calculated with the new experimentally determined field-free energies of the f and g states. Dotted lines represent calculations in which the corresponding $n$h$0_5$ state was shifted by $\pm2$~MHz with respect to the position predicted by the polarisation model. This error margin leads to a shift of up to 50~kHz in the Stark states of the manifold at higher fields (see \sref{sec:sensitivities}). The shift of the transition to the 50g$0_4$ state, depicted in \fref{fig:quadratic}a, shows excellent agreement with the calculation. In the case of 70g$0_4$, depicted in \fref{fig:quadratic}b, the fitted line centres are systematically too low and gradually approach the predicted energies for larger fields, where the influence of inhomogeneous or residual perpendicular stray fields diminishes. We consider the comparison between calculation and experimental spectra satisfactory and exclude an error of more than 2~MHz in the field-free position of the 70h$0_5$ state, considering the good agreement at $n=50$ and the fact that there are no perturbations predicted by MQDT \cite{jungenPrivate}.

\section{Conclusion and Outlook}

In this article, we have reported on a precision measurement of high-$n$ ($n=50$ and 70) Rydberg--Stark states of molecular hydrogen having an $\mathrm{X}^+\,^2\Sigma^+_{\mathrm{g}}(v^+=0,N^+=0)$ ion core by mmW spectroscopy. For comparison with the experimental data, we have calculated the positions of these states using a matrix-diagonalisation approach. In this approach, we determined the field-free positions of the low-$\ell$ states from a combination of MQDT calculations and experiments to a precision of better than 1~MHz, and evaluated the (off-diagonal) elements of the Stark Hamiltonian numerically. After carefully assessing the errors in the calculated Stark shifts resulting from uncertainties in the quantum defects, we could estimate errors in the calculated positions of the Rydberg Stark manifolds to be of the order of 50~kHz or less.

Comparison of computational results with measured mmW transitions to the Stark manifold at $n=50$ and 70 yielded the Stark shift of the initial $n^\prime$d$0_2$ state independently from each measured Stark state of the respective manifold. This enabled us to calibrate the applied electric fields and quantify the remaining stray electric fields in the excitation volume to better than 50\,$\mu$V/cm. The resulting agreement enabled us to verify that the global accuracy of the calculated positions is indeed better than 50~kHz for $n=50$ and better than 100~kHz for $n=70$. From polynomial fits of the averaged Stark shifts of the initial $n^\prime$d$0_2$ states, we obtained their field-free position relative to the zero-quantum-defect positions of the final Stark manifolds with a precision of better than 50~kHz. Through the measured network of mmW transitions, we could then also determine the absolute binding energies and effective quantum numbers of $n=50$ and 70 Rydberg states with $\ell=3$ and 4. Because of the weak sensitivity of the Stark states in the manifold to the field-free positions of states $\ket{n\ell N^+_N}$, our method allows us to determine quantum defects of low-$\ell$ states experimentally to a much higher precision than currently possible by MQDT.

The procedure to measure and calculate Stark spectra of high Rydberg states of H$_2$ introduced and validated in the present article represents a key step for the next generation of measurements of the dissociation and ionisation energies of H$_2$, which we anticipate will soon be possible at an accuracy of better than 100~kHz. This procedure is also applicable to other molecular systems.

\section*{Acknowledgements} 

We thank Prof. Christian Jungen (Orsay) for allowing us to use his MQDT program and for providing \emph{ab initio} quantum defects from a polarisation model for states with $\ell=(4,5)$, which we used for reference. 

This work is supported financially by the Swiss National Science Foundation (grant number 2000201B-200478) and by the European Research Council through the ERC advanced grant No. 743121 under the European Union's Horizon 2020 research and innovation programme.

\bibliographystyle{natbib} 
\bibliography{grpbib, starkBib}

\clearpage

\section*{Supplementary Material} 
Given are the binding energies used in the calculations, and figures for the analysis of the $n=70$ Stark manifold.

\begin{table}[h]
	\centering
	\caption{Field-free energies $\Delta E^0$ relative to the corresponding zero-quantum-defect position for states with $n=50$ and $n=70$ used in the present calculations. \label{tab:supplement}}
	\begin{tabular}{c S[table-format=6.3, retain-explicit-plus] S[table-format=6.3, retain-explicit-plus]}
		\\[-1.5ex]
		\hline\hline
		& \multicolumn{2}{c}{$\Delta E^0$ (MHz)} \\
		\cmidrule(lr){2-3}
		\multicolumn{1}{c}{$\ell$} & \multicolumn{1}{c}{$n=50$} & \multicolumn{1}{c}{$n=70$} \\
		\hline 
		\\[-1.5ex]
		0 & +5090.2 & +1774.8 \\
		1 & -252.9 & -967.9 \\
		2 & -1433.5 & -597.1 \\
		3 & -242.5 & -94.1 \\
		4 & -70.5 & -25.7 \\
		5 & -25.3 & -9.2 \\
		6 & -10.8 & -4.0 \\
		7 & -5.2 & -1.9 \\
		8 & -2.8 & -1.0 \\
		9 & -1.6 & -0.6 \\
		10 & -1.0 & -0.4 \\
		11 & -0.6 & -0.2 \\
		\hline\hline
	\end{tabular}
\end{table}

\begin{figure}[h]
	\centering
	\subfigure{\includegraphics[trim=0cm 0.9cm 0cm 0cm, clip=true, width=\linewidth]{
			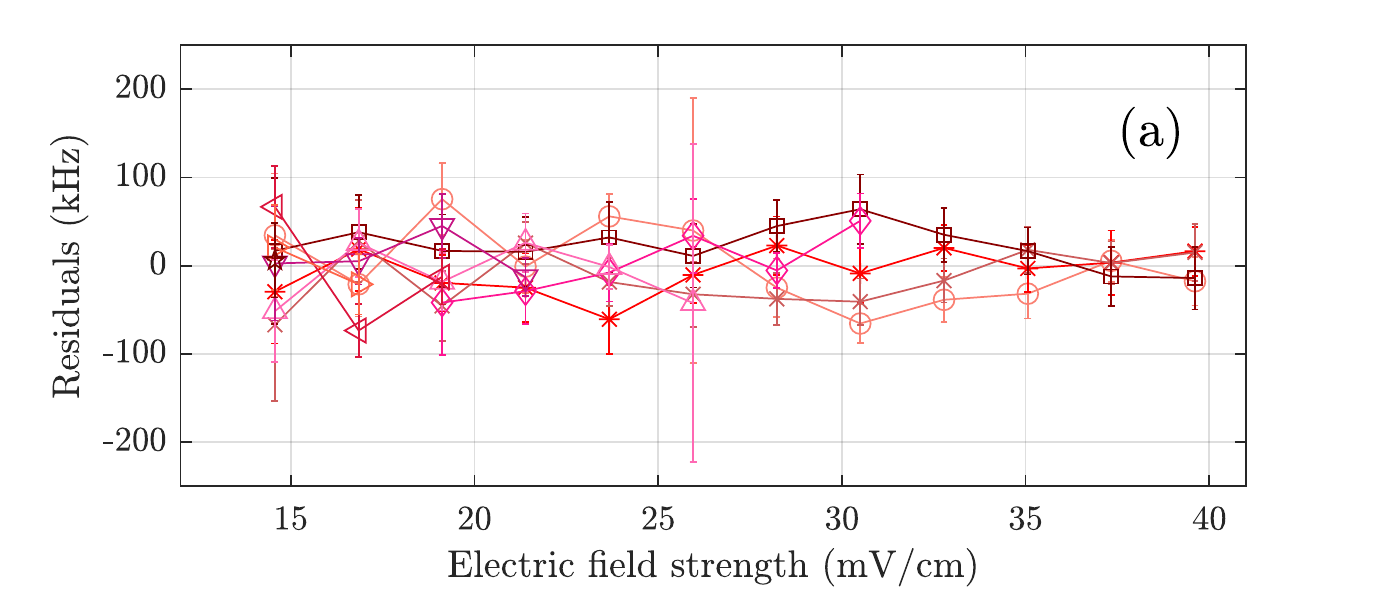}}
	\subfigure{\includegraphics[trim=0cm 0cm 0cm 0cm, clip=true, width=\linewidth]{
			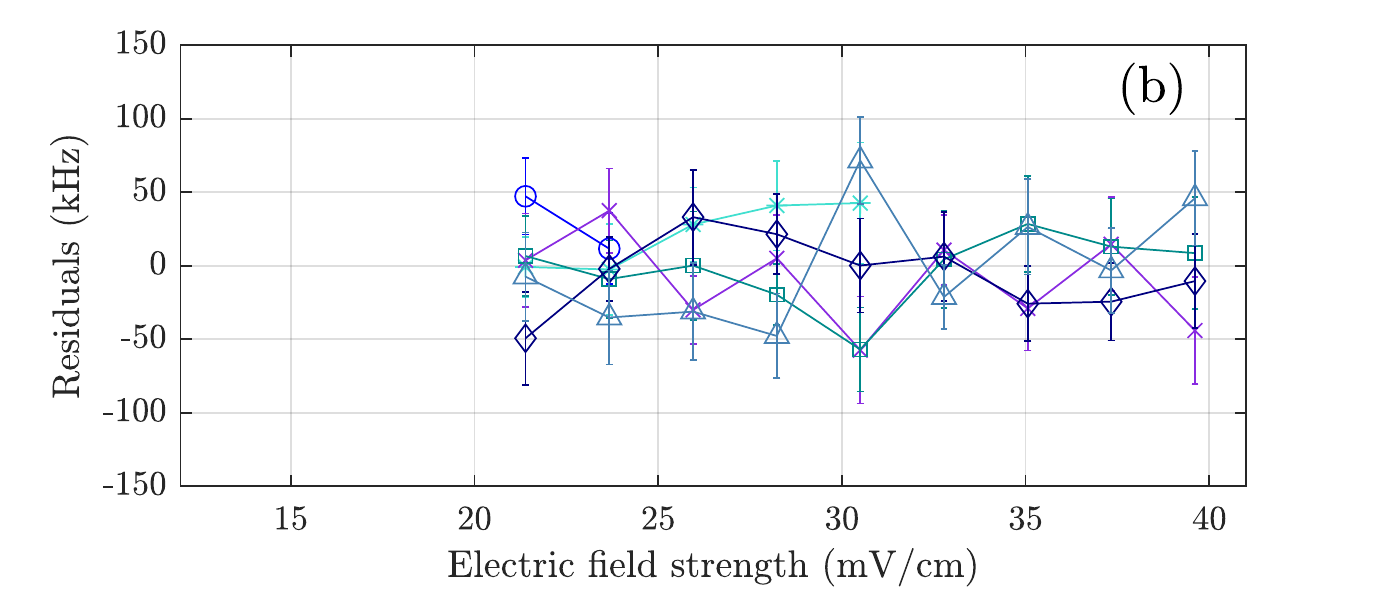}}
	\caption{Residuals of the determination of the Stark shift of the initial 63d$0_2$ state with (a) $M_N=0$ and (b) $M_N=1$. The residuals for each measured $k$ state are shown in different colours and symbols. At each field, they correspond to the deviation from the mean Stark shift of the initial state determined from all measured $k$ states.}
	\label{fig:dshift-residuals-supplement}
\end{figure}

\begin{figure}[h]
	\centering
	\subfigure{\includegraphics[trim=0cm 1.5cm 1cm 1cm, clip=true, width=\linewidth]{
		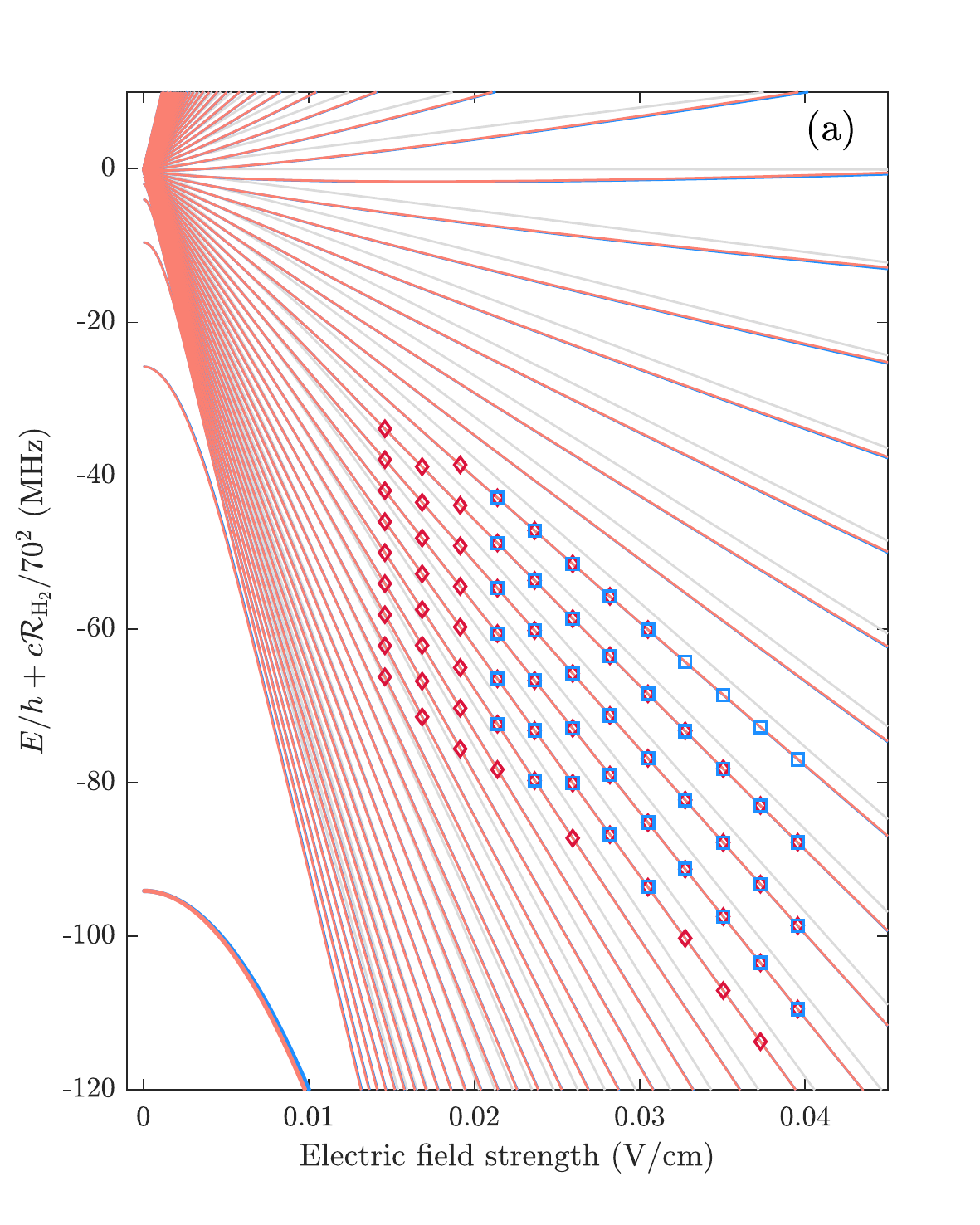}}
	\subfigure{\includegraphics[trim=0cm 0.88cm 1cm 0.19cm, clip=true, width=\linewidth]{
		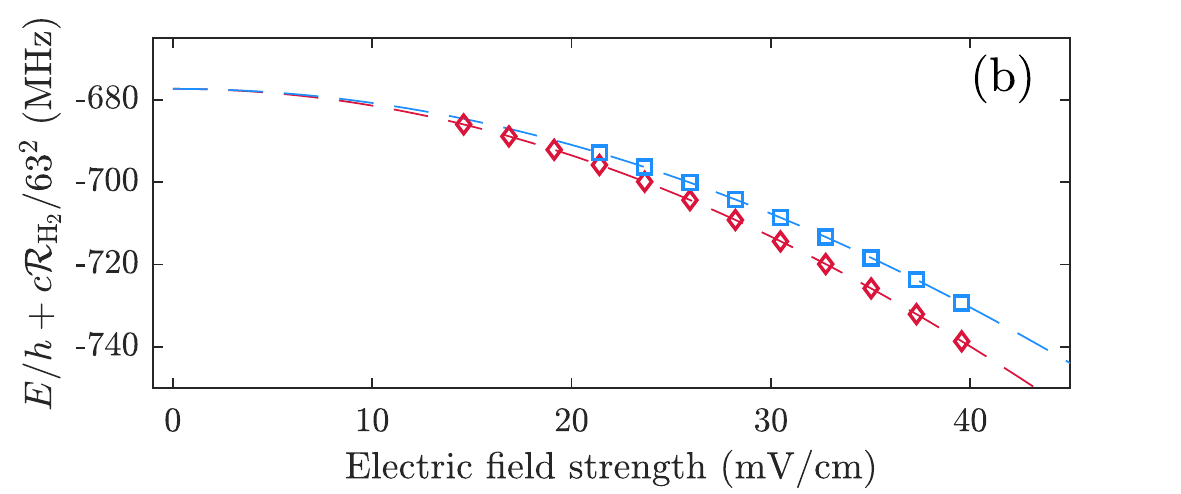}}
	\subfigure{\includegraphics[trim=0cm 0cm 1cm 0.2cm, clip=true, width=\linewidth]{
		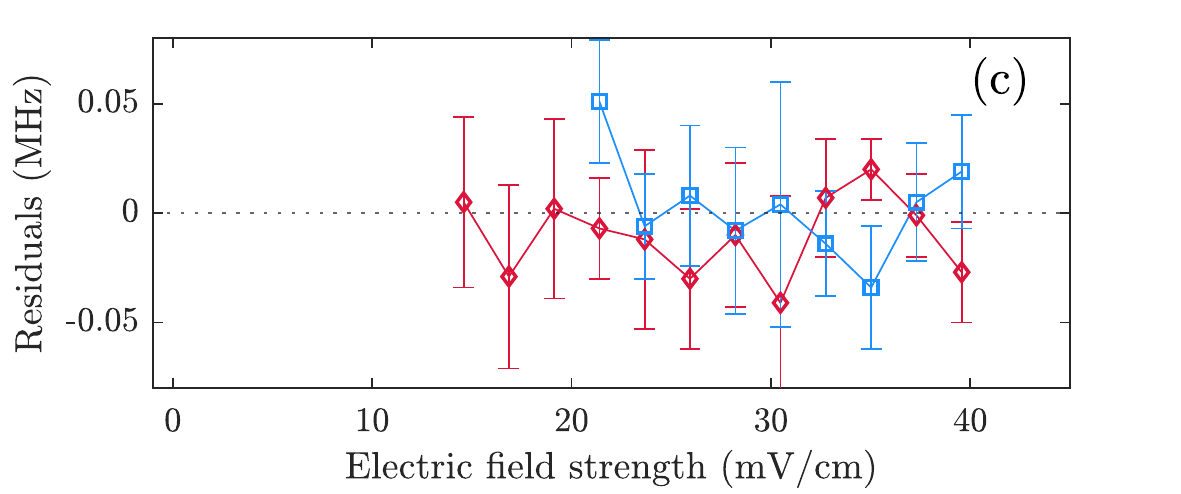}}
	\caption{(a) Calculated Stark map for $n=70$ in the region of the measured $k$ states with the origin of the energy axis set at the position of zero quantum defect. The states with $M_N=0$ and $M_N=1$ are drawn as red and blue lines, respectively. They are almost degenerate and therefore only distinguishable at the highest field. For comparison the near-hydrogenic Stark states with $M_N=11$ are drawn in light grey. Red diamonds and blue squares mark at which fields and to which $k$ states transitions were recorded (see \fref{fig:manifold-spectra}). (b) Averaged shifts of the initial states 63d$0_2(M_N=0,1)$ (open squares and diamonds) obtained by subtracting the transition frequencies from the calculated Stark-state positions. Fitted Stark shifts are drawn as dashed lines, obtained by simultaneously fitting two polynomials with the same field-free position. (c) Corresponding residuals of the combined fit.}
	\label{fig:supplement}
\end{figure}

\end{document}